\documentclass[aps,onecolumn,preprint,groupedaddress,nofootinbib,showpacs]{revtex4-1}

\usepackage{articlestyle}
\usepackage{tikz}

\usepgflibrary{fpu}

\def\EtacOneS{2.9836}
\def\EtacOneSErr{.0007}
\def\PsiOneS{3.096916}
\def\PsiOneSErr{0.000011}
\def\chicZOneP{3.41475}
\def\chicZOnePErr{0.00031}
\def\chicOOneP{3.51066}
\def\chicOOnePErr{0.00007}
\def\hcOneP{3.52538}
\def\hcOnePErr{0.00011}
\def\chicTOneP{3.55620}
\def\chicTOnePErr{0.00009}
\def\EtacTwoS{3.6394}
\def\EtacTwoSErr{0.0013}
\def\PsiTwoS{3.686109}
\def\PsiTwoSErr{0.000014} %Assymetrical error. Used the greater one
\def\PsiA{3.77315} %Psi3770	l=0,2	j = 1
\def\PsiAErr{0.00033}
\def\XA{3.87169} %X3872
\def\XAErr{0.00017}
\def\chicZTwoP{3.9184}
\def\chicZTwoPErr{0.0019}
\def\chicTTwoP{3.9272}
\def\chicTTwoPErr{0.0026}
\def\PsiB{4.039} %Psi4040 l=0,2	j = 1
\def\PsiBErr{0.001}
\def\PsiC{4.191} %Psi4160 l=0,2	j = 1
\def\PsiCErr{0.005}
\def\XB{4.251} %X4260
\def\XBErr{0.009} 
\def\XC{4.361} %X4360
\def\XCErr{0.013} 
\def\PsiD{4.421} %Psi4415 l=0,2	j = 1
\def\PsiDErr{0.004}
\def\XD{4.664} %X4660
\def\XDErr{0.012}
\def\UpsilonOneS{9.46030}
\def\UpsilonOneSErr{0.00026}
\def\chibZOneP{9.85944}
\def\chibZOnePErr{0.00042}
\def\chibOOneP{9.89278}
\def\chibOOnePErr{0.00026}
\def\hbOneP{9.8993}
\def\hbOnePErr{0.0010}
\def\chibTOneP{9.91221}
\def\chibTOnePErr{0.00026}
\def\UpsilonTwoS{10.02326}
\def\UpsilonTwoSErr{0.00031}
\def\UpsilonOneD{10.1637}
\def\UpsilonOneDErr{0.0014}
\def\chibZTwoP{10.2325}
\def\chibZTwoPErr{0.0004}
\def\chibOTwoP{10.25546}
\def\chibOTwoPErr{0.00022}
\def\chibTTwoP{10.26865}
\def\chibTTwoPErr{0.000022}
\def\UpsilonThreeS{10.3552}
\def\UpsilonThreeSErr{0.0002}
\def\chibThreeP{10.534}
\def\chibThreePErr{0.009}
\def\UpsilonFourS{10.5794}
\def\UpsilonFourSErr{0.0012}
\def\UpsilonA{10.876} %Upsilon 10860
\def\UpsilonAErr{0.011}
\def\UpsilonB{11.019} %Upsilon 11020
\def\UpsilonBErr{0.008}
\def\EtabTwoS{9.9990}

\def\EtabOneS{9.3980}
\def\EtabOneSErr{0.0032}
\pgfmathparse{\EtacOneS*1/4+\PsiOneS*3/4}\pgfmathsetmacro\InpCharmOneS{\pgfmathresult}
\pgfmathparse{\EtacOneSErr*1/4+\PsiOneSErr*3/4}\pgfmathsetmacro\InpCharmOneSErr{\pgfmathresult}
\pgfmathparse{\EtacTwoS*1/4+\PsiTwoS*3/4}\pgfmathsetmacro\InpCharmTwoS{\pgfmathresult}
\pgfmathparse{\EtacTwoSErr*1/4+\PsiTwoSErr*3/4}\pgfmathsetmacro\InpCharmTwoSErr{\pgfmathresult}
\pgfmathparse{\chicZOneP/12+3*\chicOOneP/12+3*\hcOneP/12+5*\chicTOneP/12}\pgfmathsetmacro\InpCharmOneP{\pgfmathresult}
\pgfmathparse{\chicZOnePErr/12+3*\chicOOnePErr/12+3*\hcOnePErr/12+5*\chicTOnePErr/12}\pgfmathsetmacro\InpCharmOnePErr{\pgfmathresult}
\pgfmathparse{\chicZTwoP/6+5*\chicTTwoP/6}\pgfmathsetmacro\InpCharmTwoP{\pgfmathresult}
\pgfmathparse{\chicZTwoPErr/6+5*\chicTTwoPErr/6}\pgfmathsetmacro\InpCharmTwoPErr{\pgfmathresult}
\pgfmathparse{\chibZOneP/12+3*\chibOOneP/12+3*\hbOneP/12+5*\chibTOneP/12}\pgfmathsetmacro\InpBottomOneP{\pgfmathresult}
\pgfmathparse{\chibZOnePErr/12+3*\chibOOnePErr/12+3*\hbOnePErr/12+5*\chibTOnePErr/12}\pgfmathsetmacro\InpBottomOnePErr{\pgfmathresult}

\pgfmathparse{\chibZTwoP/9+3*\chibOTwoP/9+5*\chibTTwoP/9}\pgfmathsetmacro\InpBottomTwoP{\pgfmathresult}
\pgfmathparse{\chibZTwoPErr/9+3*\chibOTwoPErr/9+3*\hbOnePErr/9+5*\chibTTwoPErr/9}\pgfmathsetmacro\InpBottomTwoPErr{\pgfmathresult}

\pgfmathparse{\EtacOneS*1/2+\PsiOneS*1/2}\pgfmathsetmacro\InpCharmOneSvT{\pgfmathresult}
\pgfmathparse{-\EtacOneS*1/2+\PsiOneS*1/2}\pgfmathsetmacro\InpCharmOneSErrvT{\pgfmathresult}
\pgfmathparse{\EtacTwoS*1/2+\PsiTwoS*1/2}\pgfmathsetmacro\InpCharmTwoSvT{\pgfmathresult}
\pgfmathparse{-\EtacTwoS*1/2+\PsiTwoS*1/2}\pgfmathsetmacro\InpCharmTwoSErrvT{\pgfmathresult}
\pgfmathparse{\chicZOneP/2+\chicTOneP/2}\pgfmathsetmacro\InpCharmOnePvT{\pgfmathresult}
\pgfmathparse{-\chicZOneP/2+\chicTOneP/2}\pgfmathsetmacro\InpCharmOnePErrvT{\pgfmathresult}
\pgfmathparse{\chicZTwoP/2+\chicTTwoP/2}\pgfmathsetmacro\InpCharmTwoPvT{\pgfmathresult}
\pgfmathparse{-\chicZTwoP/2+\chicTTwoP/2}\pgfmathsetmacro\InpCharmTwoPErrvT{\pgfmathresult}
\pgfmathparse{\EtabOneS/2+\UpsilonOneS/2}\pgfmathsetmacro\InpBottomOneSvT{\pgfmathresult} %Warning! Results using unconfirmed states
\pgfmathparse{-\EtabOneS/2+\UpsilonOneS/2}\pgfmathsetmacro\InpBottomOneSErrvT{\pgfmathresult} %Warning! Results using unconfirmed states

\pgfmathparse{\EtabTwoS/2+\UpsilonTwoS/2}\pgfmathsetmacro\InpBottomTwoSvT{\pgfmathresult} %Warning! Results using unconfirmed states
\pgfmathparse{-\EtabTwoS/2+\UpsilonTwoS/2}\pgfmathsetmacro\InpBottomTwoSErrvT{\pgfmathresult} %Warning! Results using unconfirmed states

\pgfmathparse{\chibZOneP/2+\chibTOneP/2}\pgfmathsetmacro\InpBottomOnePvT{\pgfmathresult}
\pgfmathparse{\chibTOneP/2.0-\chibZOneP/2.0}\pgfmathsetmacro\InpBottomOnePErrvT{\pgfmathresult}

\pgfmathparse{\chibZTwoP/2+\chibTTwoP/2}\pgfmathsetmacro\InpBottomTwoPvT{\pgfmathresult}
\pgfmathparse{-\chibZTwoP/2+\chibTTwoP/2}\pgfmathsetmacro\InpBottomTwoPErrvT{\pgfmathresult}

\def\CornellCharmOneSFitA{3.082285}
\def\CornellCharmTwoSFitA{3.743730}

\def\CornellCharmOnePFitA{3.515254}
\def\CornellCharmTwoPFitA{4.055723}

\pgfmathparse{\CornellCharmOneSFitA-\InpCharmOneS}\pgfmathsetmacro\DevCharmCornellOneSA{\pgfmathresult}
\pgfmathparse{\CornellCharmTwoSFitA-\InpCharmTwoS}\pgfmathsetmacro\DevCharmCornellTwoSA{\pgfmathresult}
\pgfmathparse{\CornellCharmOnePFitA-\InpCharmOneP}\pgfmathsetmacro\DevCharmCornellOnePA{\pgfmathresult}
\pgfmathparse{\CornellCharmTwoPFitA-\InpCharmTwoP}\pgfmathsetmacro\DevCharmCornellTwoPA{\pgfmathresult}

\def\CornellCharmOneSFitB{3.039590}
\def\CornellCharmTwoSFitB{3.662793}

\def\CornellCharmOnePFitB{3.448965}
\def\CornellCharmTwoPFitB{3.956973}

\pgfmathparse{\CornellCharmOneSFitB-\InpCharmOneS}\pgfmathsetmacro\DevCharmCornellOneSB{\pgfmathresult}
\pgfmathparse{\CornellCharmTwoSFitB-\InpCharmTwoS}\pgfmathsetmacro\DevCharmCornellTwoSB{\pgfmathresult}
\pgfmathparse{\CornellCharmOnePFitB-\InpCharmOneP}\pgfmathsetmacro\DevCharmCornellOnePB{\pgfmathresult}
\pgfmathparse{\CornellCharmTwoPFitB-\InpCharmTwoP}\pgfmathsetmacro\DevCharmCornellTwoPB{\pgfmathresult}

\def\LGPCharmOneSFitA{3.071309}
\def\LGPCharmTwoSFitA{3.721504}

\def\LGPCharmOnePFitA{3.524121}
\def\LGPCharmTwoPFitA{4.022949}

\pgfmathparse{\LGPCharmOneSFitA-\InpCharmOneS}\pgfmathsetmacro\DevCharmLGPOneSA{\pgfmathresult}
\pgfmathparse{\LGPCharmTwoSFitA-\InpCharmTwoS}\pgfmathsetmacro\DevCharmLGPTwoSA{\pgfmathresult}
\pgfmathparse{\LGPCharmOnePFitA-\InpCharmOneP}\pgfmathsetmacro\DevCharmLGPOnePA{\pgfmathresult}
\pgfmathparse{\LGPCharmTwoPFitA-\InpCharmTwoP}\pgfmathsetmacro\DevCharmLGPTwoPA{\pgfmathresult}

\def\LGPCharmOneSFitB{3.033730}
\def\LGPCharmTwoSFitB{3.664043}

\def\LGPCharmOnePFitB{3.473965}
\def\LGPCharmTwoPFitB{3.956660}

\pgfmathparse{\LGPCharmOneSFitB-\InpCharmOneS}\pgfmathsetmacro\DevCharmLGPOneSB{\pgfmathresult}
\pgfmathparse{\LGPCharmTwoSFitB-\InpCharmTwoS}\pgfmathsetmacro\DevCharmLGPTwoSB{\pgfmathresult}
\pgfmathparse{\LGPCharmOnePFitB-\InpCharmOneP}\pgfmathsetmacro\DevCharmLGPOnePB{\pgfmathresult}
\pgfmathparse{\LGPCharmTwoPFitB-\InpCharmTwoP}\pgfmathsetmacro\DevCharmLGPTwoPB{\pgfmathresult}

\def\CornellBottomOneSFitA{9.468027}
\def\CornellBottomTwoSFitA{10.042051}
\def\CornellBottomThreeSFitA{10.472754}
\def\CornellBottomFourSFitA{10.842480}

\def\CornellBottomOnePFitA{9.862324}
\def\CornellBottomTwoPFitA{10.314746}
\def\CornellBottomThreePFitA{10.697949}

\def\CornellBottomOneDFitA{10.145684}

\pgfmathparse{\CornellBottomOneSFitA-\UpsilonOneS}\pgfmathsetmacro\DevBottomCornellOneSA{\pgfmathresult}
\pgfmathparse{\CornellBottomTwoSFitA-\UpsilonTwoS}\pgfmathsetmacro\DevBottomCornellTwoSA{\pgfmathresult}
\pgfmathparse{\CornellBottomThreeSFitA-\UpsilonThreeS}\pgfmathsetmacro\DevBottomCornellThreeSA{\pgfmathresult}
\pgfmathparse{\CornellBottomFourSFitA-\UpsilonFourS}\pgfmathsetmacro\DevBottomCornellFourSA{\pgfmathresult}
\pgfmathparse{\CornellBottomOnePFitA-\InpBottomOneP}\pgfmathsetmacro\DevBottomCornellOnePA{\pgfmathresult}
\pgfmathparse{\CornellBottomTwoPFitA-\InpBottomTwoP}\pgfmathsetmacro\DevBottomCornellTwoPA{\pgfmathresult}
\pgfmathparse{\CornellBottomThreePFitA-\chibThreeP}\pgfmathsetmacro\DevBottomCornellThreePA{\pgfmathresult}
\pgfmathparse{\CornellBottomOneDFitA-\UpsilonOneD}\pgfmathsetmacro\DevBottomCornellOneDA{\pgfmathresult}

\def\CornellBottomOneSFitB{9.456504}
\def\CornellBottomTwoSFitB{10.019355}
\def\CornellBottomThreeSFitB{10.441152}
\def\CornellBottomFourSFitB{10.803105}

\def\CornellBottomOnePFitB{9.843770}
\def\CornellBottomTwoPFitB{10.286777}
\def\CornellBottomThreePFitB{10.661895}

\def\CornellBottomOneDFitB{10.121426}

\pgfmathparse{\CornellBottomOneSFitB-\UpsilonOneS}\pgfmathsetmacro\DevBottomCornellOneSB{\pgfmathresult}
\pgfmathparse{\CornellBottomTwoSFitB-\UpsilonTwoS}\pgfmathsetmacro\DevBottomCornellTwoSB{\pgfmathresult}
\pgfmathparse{\CornellBottomThreeSFitB-\UpsilonThreeS}\pgfmathsetmacro\DevBottomCornellThreeSB{\pgfmathresult}
\pgfmathparse{\CornellBottomFourSFitB-\UpsilonFourS}\pgfmathsetmacro\DevBottomCornellFourSB{\pgfmathresult}
\pgfmathparse{\CornellBottomOnePFitB-\InpBottomOnePvT}\pgfmathsetmacro\DevBottomCornellOnePB{\pgfmathresult}
\pgfmathparse{\CornellBottomTwoPFitB-\InpBottomTwoPvT}\pgfmathsetmacro\DevBottomCornellTwoPB{\pgfmathresult}
\pgfmathparse{\CornellBottomThreePFitB-\chibThreeP}\pgfmathsetmacro\DevBottomCornellThreePB{\pgfmathresult}
\pgfmathparse{\CornellBottomOneDFitB-\UpsilonOneD}\pgfmathsetmacro\DevBottomCornellOneDB{\pgfmathresult}

\def\LGPBottomOneSFitA{9.465996}
\def\LGPBottomTwoSFitA{10.036895}
\def\LGPBottomThreeSFitA{10.432168}
\def\LGPBottomFourSFitA{10.763105}

\def\LGPBottomOnePFitA{9.874473}
\def\LGPBottomTwoPFitA{10.294629}
\def\LGPBottomThreePFitA{10.638301}

\def\LGPBottomOneDFitA{10.150176}

\pgfmathparse{\LGPBottomOneSFitA-\UpsilonOneS}\pgfmathsetmacro\DevBottomLGPOneSA{\pgfmathresult}
\pgfmathparse{\LGPBottomTwoSFitA-\UpsilonTwoS}\pgfmathsetmacro\DevBottomLGPTwoSA{\pgfmathresult}
\pgfmathparse{\LGPBottomThreeSFitA-\UpsilonThreeS}\pgfmathsetmacro\DevBottomLGPThreeSA{\pgfmathresult}
\pgfmathparse{\LGPBottomFourSFitA-\UpsilonFourS}\pgfmathsetmacro\DevBottomLGPFourSA{\pgfmathresult}
\pgfmathparse{\LGPBottomOnePFitA-\InpBottomOneP}\pgfmathsetmacro\DevBottomLGPOnePA{\pgfmathresult}
\pgfmathparse{\LGPBottomTwoPFitA-\InpBottomTwoP}\pgfmathsetmacro\DevBottomLGPTwoPA{\pgfmathresult}
\pgfmathparse{\LGPBottomThreePFitA-\chibThreeP}\pgfmathsetmacro\DevBottomLGPThreePA{\pgfmathresult}
\pgfmathparse{\LGPBottomOneDFitA-\UpsilonOneD}\pgfmathsetmacro\DevBottomLGPOneDA{\pgfmathresult}

\def\LGPBottomOneSFitB{9.454824}
\def\LGPBottomTwoSFitB{10.013652}
\def\LGPBottomThreeSFitB{10.399121}
\def\LGPBottomFourSFitB{10.721621}

\def\LGPBottomOnePFitB{9.855801}
\def\LGPBottomTwoPFitB{10.265527}
\def\LGPBottomThreePFitB{10.600371}

\def\LGPBottomOneDFitB{10.125254}

\pgfmathparse{\LGPBottomOneSFitB-\UpsilonOneS}\pgfmathsetmacro\DevBottomLGPOneSB{\pgfmathresult}
\pgfmathparse{\LGPBottomTwoSFitB-\UpsilonTwoS}\pgfmathsetmacro\DevBottomLGPTwoSB{\pgfmathresult}
\pgfmathparse{\LGPBottomThreeSFitB-\UpsilonThreeS}\pgfmathsetmacro\DevBottomLGPThreeSB{\pgfmathresult}
\pgfmathparse{\LGPBottomFourSFitB-\UpsilonFourS}\pgfmathsetmacro\DevBottomLGPFourSB{\pgfmathresult}
\pgfmathparse{\LGPBottomOnePFitB-\InpBottomOnePvT}\pgfmathsetmacro\DevBottomLGPOnePB{\pgfmathresult}
\pgfmathparse{\LGPBottomTwoPFitB-\InpBottomTwoPvT}\pgfmathsetmacro\DevBottomLGPTwoPB{\pgfmathresult}
\pgfmathparse{\LGPBottomThreePFitB-\chibThreeP}\pgfmathsetmacro\DevBottomLGPThreePB{\pgfmathresult}
\pgfmathparse{\LGPBottomOneDFitB-\UpsilonOneD}\pgfmathsetmacro\DevBottomLGPOneDB{\pgfmathresult}

\def\LGPBottomOneSFitC{9.446777}
\def\LGPBottomTwoSFitC{10.017793}
\def\LGPBottomThreeSFitC{10.413262}
\def\LGPBottomFourSFitC{10.744395}

\def\LGPBottomOnePFitC{9.855254}
\def\LGPBottomTwoPFitC{10.275605}
\def\LGPBottomThreePFitC{10.619512}

\def\LGPBottomOneDFitC{10.131074}

\pgfmathparse{\LGPBottomOneSFitC-\InpBottomOneSvT}\pgfmathsetmacro\DevBottomLGPOneSC{\pgfmathresult}
\pgfmathparse{\LGPBottomTwoSFitC-\InpBottomTwoSvT}\pgfmathsetmacro\DevBottomLGPTwoSC{\pgfmathresult}
\pgfmathparse{\LGPBottomThreeSFitC-\UpsilonThreeS}\pgfmathsetmacro\DevBottomLGPThreeSC{\pgfmathresult}
\pgfmathparse{\LGPBottomFourSFitC-\UpsilonFourS}\pgfmathsetmacro\DevBottomLGPFourSC{\pgfmathresult}
\pgfmathparse{\LGPBottomOnePFitC-\InpBottomOnePvT}\pgfmathsetmacro\DevBottomLGPOnePC{\pgfmathresult}
\pgfmathparse{\LGPBottomTwoPFitC-\InpBottomTwoPvT}\pgfmathsetmacro\DevBottomLGPTwoPC{\pgfmathresult}
\pgfmathparse{\LGPBottomThreePFitC-\chibThreeP}\pgfmathsetmacro\DevBottomLGPThreePC{\pgfmathresult}
\pgfmathparse{\LGPBottomOneDFitC-\UpsilonOneD}\pgfmathsetmacro\DevBottomLGPOneDC{\pgfmathresult}

\def\CornellBottomOneSFitC{9.456504}
\def\CornellBottomTwoSFitC{10.019355}
\def\CornellBottomThreeSFitC{10.441152}
\def\CornellBottomFourSFitC{10.803105}

\def\CornellBottomOnePFitC{9.843770}
\def\CornellBottomTwoPFitC{10.286777}
\def\CornellBottomThreePFitC{10.661895}

\def\CornellBottomOneDFitC{10.121426}

\pgfmathparse{\CornellBottomOneSFitC-\InpBottomOneSvT}\pgfmathsetmacro\DevBottomCornellOneSC{\pgfmathresult}
\pgfmathparse{\CornellBottomTwoSFitC-\InpBottomTwoSvT}\pgfmathsetmacro\DevBottomCornellTwoSC{\pgfmathresult}
\pgfmathparse{\CornellBottomThreeSFitC-\UpsilonThreeS}\pgfmathsetmacro\DevBottomCornellThreeSC{\pgfmathresult}
\pgfmathparse{\CornellBottomFourSFitC-\UpsilonFourS}\pgfmathsetmacro\DevBottomCornellFourSC{\pgfmathresult}
\pgfmathparse{\CornellBottomOnePFitC-\InpBottomOnePvT}\pgfmathsetmacro\DevBottomCornellOnePC{\pgfmathresult}
\pgfmathparse{\CornellBottomTwoPFitC-\InpBottomTwoPvT}\pgfmathsetmacro\DevBottomCornellTwoPC{\pgfmathresult}
\pgfmathparse{\CornellBottomThreePFitC-\chibThreeP}\pgfmathsetmacro\DevBottomCornellThreePC{\pgfmathresult}
\pgfmathparse{\CornellBottomOneDFitC-\UpsilonOneD}\pgfmathsetmacro\DevBottomCornellOneDC{\pgfmathresult}

\def\CornellCharmOneSFitSimFit{2.930020}
\def\CornellCharmTwoSFitSimFit{3.687441}

\def\CornellCharmOnePFitSimFit{3.419082}
\def\CornellCharmTwoPFitSimFit{4.043652}

\pgfmathparse{\CornellCharmOneSFitSimFit-\InpCharmOneS}\pgfmathsetmacro\DevCharmCornellOneSSimFit{\pgfmathresult}
\pgfmathparse{\CornellCharmTwoSFitSimFit-\InpCharmTwoS}\pgfmathsetmacro\DevCharmCornellTwoSSimFit{\pgfmathresult}
\pgfmathparse{\CornellCharmOnePFitSimFit-\InpCharmOneP}\pgfmathsetmacro\DevCharmCornellOnePSimFit{\pgfmathresult}
\pgfmathparse{\CornellCharmTwoPFitSimFit-\InpCharmTwoP}\pgfmathsetmacro\DevCharmCornellTwoPSimFit{\pgfmathresult}

\def\LGPCharmOneSFitSimFit{2.964160}
\def\LGPCharmTwoSFitSimFit{3.685293}

\def\LGPCharmOnePFitSimFit{3.458418}
\def\LGPCharmTwoPFitSimFit{4.019941}

\pgfmathparse{\LGPCharmOneSFitSimFit-\InpCharmOneS}\pgfmathsetmacro\DevCharmLGPOneSSimFit{\pgfmathresult}
\pgfmathparse{\LGPCharmTwoSFitSimFit-\InpCharmTwoS}\pgfmathsetmacro\DevCharmLGPTwoSSimFit{\pgfmathresult}
\pgfmathparse{\LGPCharmOnePFitSimFit-\InpCharmOneP}\pgfmathsetmacro\DevCharmLGPOnePSimFit{\pgfmathresult}
\pgfmathparse{\LGPCharmTwoPFitSimFit-\InpCharmTwoP}\pgfmathsetmacro\DevCharmLGPTwoPSimFit{\pgfmathresult}

\def\CornellBottomOneSFitSimFit{9.486738}
\def\CornellBottomTwoSFitSimFit{10.002598}
\def\CornellBottomThreeSFitSimFit{10.386855}
\def\CornellBottomFourSFitSimFit{10.715996}

\def\CornellBottomOnePFitSimFit{9.844629}
\def\CornellBottomTwoPFitSimFit{10.247871}
\def\CornellBottomThreePFitSimFit{10.588770}

\def\CornellBottomOneDFitSimFit{10.098340}

\pgfmathparse{\CornellBottomOneSFitSimFit-\InpBottomOneSvT}\pgfmathsetmacro\DevBottomCornellOneSSimFit{\pgfmathresult}
\pgfmathparse{\CornellBottomTwoSFitSimFit-\InpBottomTwoSvT}\pgfmathsetmacro\DevBottomCornellTwoSSimFit{\pgfmathresult}
\pgfmathparse{\CornellBottomThreeSFitSimFit-\UpsilonThreeS}\pgfmathsetmacro\DevBottomCornellThreeSSimFit{\pgfmathresult}
\pgfmathparse{\CornellBottomFourSFitSimFit-\UpsilonFourS}\pgfmathsetmacro\DevBottomCornellFourSSimFit{\pgfmathresult}
\pgfmathparse{\CornellBottomOnePFitSimFit-\InpBottomOnePvT}\pgfmathsetmacro\DevBottomCornellOnePSimFit{\pgfmathresult}
\pgfmathparse{\CornellBottomTwoPFitSimFit-\InpBottomTwoPvT}\pgfmathsetmacro\DevBottomCornellTwoPSimFit{\pgfmathresult}
\pgfmathparse{\CornellBottomThreePFitSimFit-\chibThreeP}\pgfmathsetmacro\DevBottomCornellThreePSimFit{\pgfmathresult}
\pgfmathparse{\CornellBottomOneDFitSimFit-\UpsilonOneD}\pgfmathsetmacro\DevBottomCornellOneDSimFit{\pgfmathresult}

\def\LGPBottomOneSFitSimFit{9.470762}
\def\LGPBottomTwoSFitSimFit{10.004629}
\def\LGPBottomThreeSFitSimFit{10.369668}
\def\LGPBottomFourSFitSimFit{10.674590}

\def\LGPBottomOnePFitSimFit{9.856387}
\def\LGPBottomTwoPFitSimFit{10.244395}
\def\LGPBottomThreePFitSimFit{10.560801}

\def\LGPBottomOneDFitSimFit{10.112832}

\pgfmathparse{\LGPBottomOneSFitSimFit-\InpBottomOneSvT}\pgfmathsetmacro\DevBottomLGPOneSSimFit{\pgfmathresult}
\pgfmathparse{\LGPBottomTwoSFitSimFit-\InpBottomTwoSvT}\pgfmathsetmacro\DevBottomLGPTwoSSimFit{\pgfmathresult}
\pgfmathparse{\LGPBottomThreeSFitSimFit-\UpsilonThreeS}\pgfmathsetmacro\DevBottomLGPThreeSSimFit{\pgfmathresult}
\pgfmathparse{\LGPBottomFourSFitSimFit-\UpsilonFourS}\pgfmathsetmacro\DevBottomLGPFourSSimFit{\pgfmathresult}
\pgfmathparse{\LGPBottomOnePFitSimFit-\InpBottomOnePvT}\pgfmathsetmacro\DevBottomLGPOnePSimFit{\pgfmathresult}
\pgfmathparse{\LGPBottomTwoPFitSimFit-\InpBottomTwoPvT}\pgfmathsetmacro\DevBottomLGPTwoPSimFit{\pgfmathresult}
\pgfmathparse{\LGPBottomThreePFitSimFit-\chibThreeP}\pgfmathsetmacro\DevBottomLGPThreePSimFit{\pgfmathresult}
\pgfmathparse{\LGPBottomOneDFitSimFit-\UpsilonOneD}\pgfmathsetmacro\DevBottomLGPOneDSimFit{\pgfmathresult}

\usepackage{dcolumn}% Align table columns on decimal point

\begin{document}

\preprint{LPT-Orsay-17-13}

\title{Lattice Gluon Propagator and One-Gluon-Exchange Potential}

\author{Attilio Cucchieri}
\email[]{attilio@ifsc.usp.br}
\affiliation{Instituto de F\'isica de S\~ao Carlos, Universidade de S\~ao Paulo \\
Caixa Postal 369, CEP 13560-970, S\~ao Carlos SP, Brazil}
\affiliation{Laboratoire de Physique Th\'eorique, CNRS, Univ.\ Paris-Sud et Universit\'e Paris-Saclay, B\^atiment 210, 91405 Orsay Cedex, France}
\author{Tereza Mendes}
\email[]{mendes@ifsc.usp.br}
\affiliation{Instituto de F\'isica de S\~ao Carlos, Universidade de S\~ao Paulo \\
Caixa Postal 369, CEP 13560-970, S\~ao Carlos SP, Brazil}
\author{Willian M.\ Serenone}
\email[]{willian.serenone@usp.br}
\affiliation{Instituto de F\'isica de S\~ao Carlos, Universidade de S\~ao Paulo \\
Caixa Postal 369, CEP 13560-970, S\~ao Carlos SP, Brazil}

\date{\today}

\begin{abstract}
We consider the interquark potential in the one-gluon-exchange 
(OGE) approximation, using a fully nonperturbative gluon propagator
from large-volume lattice simulations.
The resulting $V_{LGP}$ potential is non-confining, showing that the
OGE approximation is not sufficient to describe the infrared sector
of QCD. Nevertheless, it represents an improvement over the perturbative 
(Coulomb-like) potential, 
since it allows the description of a few low-lying bound states of
charmonium and bottomonium. 
In order to achieve a better description of these spectra, we add
to $V_{LGP}$ a linearly growing term. The obtained results are
comparable to the corresponding ones in the Cornell-potential case. 
As a byproduct of our study, we estimate the interquark distance for
the considered charmonium and bottomonium states.
\end{abstract}


\maketitle

\section{Introduction}
\label{sec:Introduction}

A reliable description of heavy quarkonia states is of great interest for our understanding of 
nonperturbative aspects of QCD \cite{Brambilla:2014jmp} and is expected to be important 
in guiding the search for physics beyond the standard model \cite{Love:2008ys}.
A fortuitous advantage in the study of such states is that, due to the large mass of the 
heavy quarks, various approximations may be adopted. For example, an expansion in inverse 
powers of the heavy-quark mass $m$ is performed in potential nonrelativistic QCD (pNRQCD) 
\cite{Brambilla:2004jw}, and lattice simulations (especially for bottomonium systems) are 
applied to effective actions obtained by an expansion in powers of the heavy-quark velocity $v/c$.
Similarly, in the relativistic quark model with the quasipotential approach, radiative corrections may
be included and treated perturbatively in the case of heavy quarkonia \cite{Ebert:2011jc}.
This possibility of exploring different scales of the problem separately is also
helpful in methods more directly based on QCD, such as studies of Dyson-Schwinger and Bethe-Salpeter
equations \cite{Fischer:2014cfa}. 

An early but still successful approach to describe heavy quarkonia is given by nonrelativistic 
potential models, to which relativistic corrections may also be added \cite{Radford:2007vd}.\footnote{Note that these corrections may be computed from lattice data for the Wilson loop \cite{Koma:2006si,Koma:2008zza}.}
The idea is to view confinement as an ``a priori'' property of QCD, modeling
the interquark potential to incorporate some known features of the interaction at both
ends of the energy scale, i.e.\ at small and large distances.
The simplest such model, the Cornell --- or Coulomb-plus-linear --- potential
\cite{Eichten:1978tg,Eichten:1979ms,Eichten:2002qv}, is 
obtained by supplementing the high-energy (perturbative) part of the potential 
with an explicit 
confining term. Hence, the resulting expression is a sum of two terms: the first 
one comes from the 
quark-antiquark interaction in the one-gluon-exchange (OGE) approximation  
using a tree-level gluon propagator,
and the second one is a linearly rising potential.
We have
\begin{equation}
\label{Cornell}
V(r) = -\frac{4}{3}\frac{\alpha_s}{r} \,+\, \sigma\,r\,,
\end{equation}
where ${\alpha_s}$ is the strong coupling constant and $\sigma$ is the string tension.
The first term may be associated with scattering of the quark-antiquark pair inside the meson and
is analogous to the Coulomb potential in the QED case. The second term corresponds to linear 
confinement as observed from the strong-coupling expansion of the Wilson loop in lattice gauge theory
with static quarks.
The Cornell potential provides a spin-independent description of
the interquark potential for heavy quarks, with parameters determined by fitting a few 
known states (see e.g.\ \cite{Chung:2008sm}) or by comparison with lattice simulations.
For a recent determination of these parameters, see Ref.\ \cite{Kawanai:2011xb}.
The numerical procedure for obtaining the mass spectrum for the 
Cornell potential, as well as for other commonly used potentials is reviewed in 
detail in \cite{Lucha:1991vn}.

More generally, the static interquark potential may be defined conveniently in terms of the
Wilson loop, or it may be obtained (perturbatively) by taking the nonrelativistic limit in the Bethe-Salpeter 
equation describing the bound state of two heavy fermions. This yields a Schr\"odinger equation, to which
a linear term is added a posteriori. 
It would be interesting, nevertheless, to have a better insight about confinement as an {\em emergent} property of the interquark interaction induced by the gluon propagator, rather than as a built-in feature.
Of course, at the hadronic scale, 
the full gluon propagator in QCD is very different from the perturbative one
and it should contain information about confinement.
In order to use this nonperturbative information we propose to substitute the free gluon 
propagator in the OGE term of the potential, as described above, by a fully nonperturbative one, 
obtained from lattice simulations. We want to check if this replacement leads to an improved 
description of the spectra, possibly without the need to include the linearly rising term explicitly.
To this end, we use the data generated in studies of the SU(2) gluon
propagator in Landau gauge on very large lattices (up to $128^4$), reported in
\cite{Cucchieri:2007md,Cucchieri:2007rg}.
We note that lattice data for propagators in the $SU(2)$ and $SU(3)$
cases have essentially the same behavior apart from a global constant
\cite{Cucchieri:2007zm}, which can be fixed by choosing a specific
multiplicative renormalization condition, as done in the momentum-subtraction
scheme.
Of course, to include all QCD effects in the analysis, one should consider
a gluon propagator obtained from unquenched SU(3) simulations. 
On the other hand, such simulations have been done 
\cite{Bowman:2004jm,Ilgenfritz:2006he,Kamleh:2007ud,Silva:2010vx}
only for rather small physical volumes up to now, and with associated
unquenching effects that seem to be modest, at the quantitative rather 
than qualitative level. Moreover, we are interested in the origin of 
the linearly confining term of the static interquark potential, 
which should already show up in the pure-Yang-Mills sector of the
theory, which is confining, without the need to include unquenching
effects. Thus, we choose to use our SU(2) lattice data \cite{Cucchieri:2007md,
Cucchieri:2007rg,Cucchieri:2007zm}, for which data with good accuracy 
and well controlled finite-volume effects are available.

We organize this paper in the following way. 
In Section \ref{sec:Pot_Model_Review} we review
the procedure for obtaining the Coulomb potential in QED as 
the nonrelativistic limit of $e^- e^+$ 
scattering (at tree level) and the analogous calculation in the
heavy-quark case. 
We then follow the same procedure using
the lattice gluon propagator to obtain a nonperturbatively
corrected OGE potential, i.e.\
we use directly the fit obtained in Ref.\ \cite{Cucchieri:2011ig}
and perform the Fourier transform analytically to get the potential.
The result is compared to the perturbative (Coulomb-like) potential in
Fig.\ \ref{fig:Comparison_Potentials}.
In Section \ref{sec:Methods} we describe the numerical method for obtaining 
the mass spectra associated with a given interquark potential in the 
nonrelativistic approximation. We also outline
our choices for the interquark potentials, the fitting
parameters, and the experimental data used for input and comparison.
Our results for the spectra and interquark distances are reported in 
Section \ref{sec:Results} and our conclusions in Section \ref{sec:Conclusions}.
Preliminary versions of our study have been presented in 
Refs.\ \cite{Serenone:2012yta,Serenone:2014ota,Serenone:2015qra}.
We note again that our aim is to gain a qualitative understanding
of the interplay between perturbative
and nonperturbative features of the interquark potential.
Our approach is similar in spirit to the one in Refs.\
\cite{Gonzalez:2011zc,Vento:2012wp,Gonzalez:2012hx},
but our conclusions are somewhat different.

\section{Potential from Lattice Propagator}
\label{sec:Pot_Model_Review}

\begin{figure}
\includegraphics[width=0.6\columnwidth]{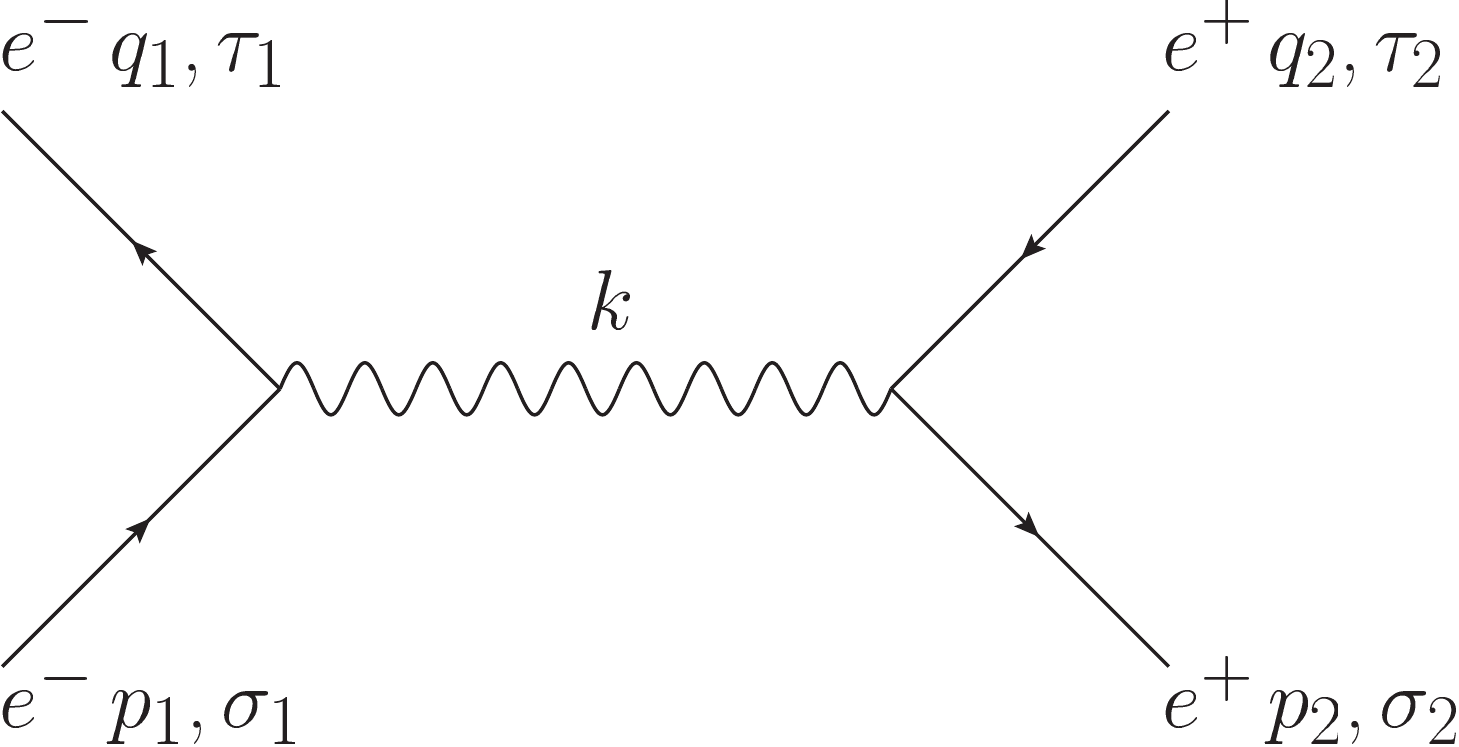}
\hfill
\includegraphics[width=0.35\columnwidth]{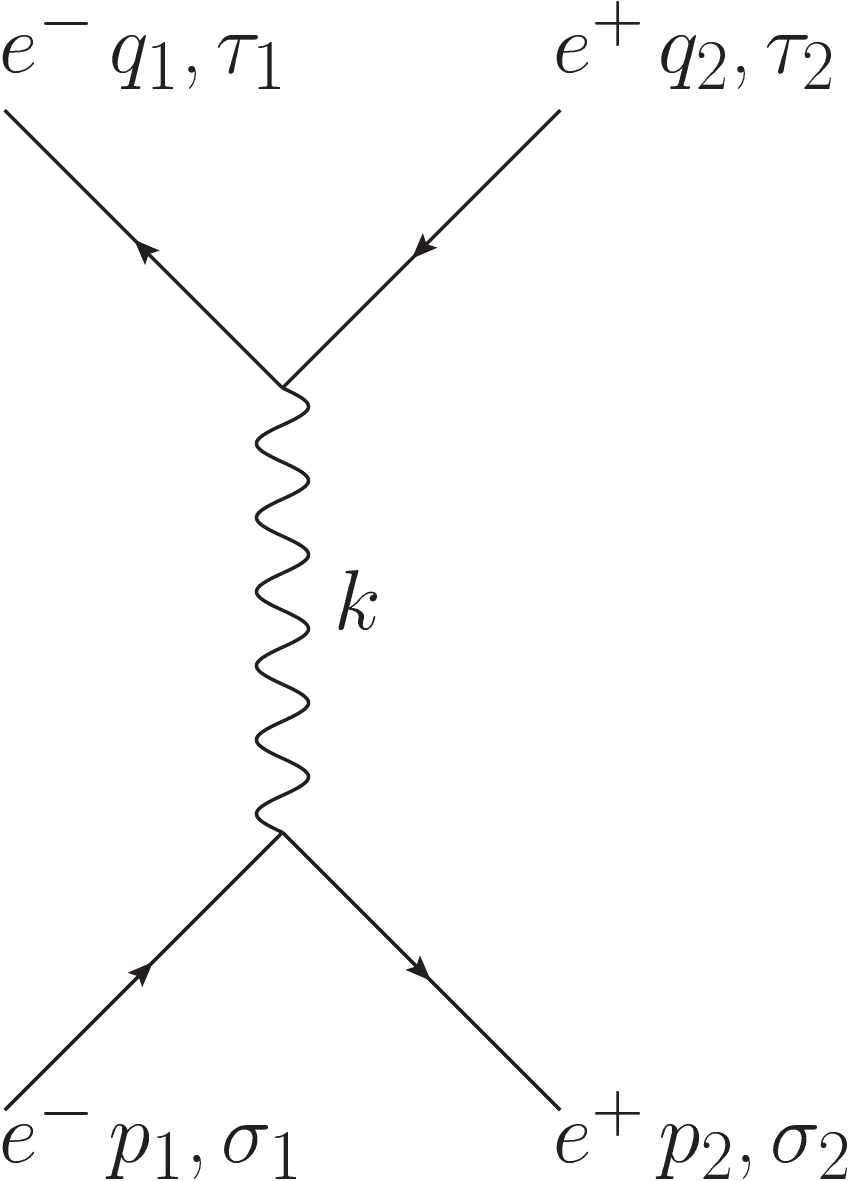}
\caption{Feynman diagrams corresponding to the two terms in the $e^- e^+$ scattering amplitude. The 
left diagram corresponds to the $t$ channel (photon exchange) and the right 
diagram to the $s$ channel (pair annihilation).}
\label{fig:feynman_graphs}
\end{figure}

Let us first review how the Coulomb potential is obtained in the nonrelativistic
limit of QED from the application of Feynman rules to the electron-positron system. The 
scattering-matrix $S_{fi}$, from which the interaction potential may be obtained, is given by 
\begin{equation}
 \label{eq:S-matrix}
 S_{fi} \;\equiv\; \langle f | i \rangle = \delta_{fi} \,+\, 
i (2 \pi)^4 \, \delta^{(4)}(Q-P)\, T_{fi} \,,
\end{equation}
where $Q$ and $P$ correspond respectively to the final and initial total
momentum and $T_{fi}$ is the scattering amplitude. The two tree-level Feynman
diagrams contributing to $T_{fi}$ (see Fig.\ \ref{fig:feynman_graphs}) correspond to 
the $t$ and $s$ channels,
respectively coming from scattering with one photon exchange and to 
annihilation and creation of an $e^- e^+$ pair. We get
\begin{equation}
\label{eq:QED_Scattering_Amplitude}
 T_{fi} = \frac{1}{(2 \pi)^6} \frac{m^2}{\sqrt{E_{p_1}E_{p_2}E_{q_1}E_{q_2}}} 
 \left(t_{\text{exch}}+t_{\text{annihil}} \right)\,,
\end{equation}
where
\begin{align}
 t_{\text{exch}} \;=&\; e^2\,\overline{u}(q_1,\tau_1)\,\gamma^\mu 
  \,u(p_1,\sigma_1)\;P_{\mu \nu}(k)\; \nonumber \\
  \quad &  \times \,\overline{v}(p_2,\sigma_2)\,\gamma^\nu\,v(q_2,\tau_2) 
\end{align}
and
\begin{align}
 t_{\text{annihil}} \;={}&\;-e^2\,\overline{v}(p_2,\sigma_2)\,\gamma^\mu
  \,u(p_1,\sigma_1)\;P_{\mu \nu}(k)\; \nonumber \\
  \quad & \times \, \overline{u}  (q_1,\tau_1)\,\gamma^\nu v(q_2,\tau_2) \;.
\end{align}
We follow the notation in \cite{bjorken1965relativistic, Lucha:1991vn, Lucha:1995zv}:
$p_i$ denotes the momentum of the incoming particles and $q_i$ of the outgoing ones.
The particles' initial and final spins are respectively $\sigma_i$ and $\tau_i$.
We represent the photon propagator by a function $P_{\mu \nu}(k)$ of the photon momentum $k$.

We then make the nonrelativistic approximation, i.e.\ we impose the kinetic 
energy of the system to be much smaller than its rest energy 
($ |\vec{p}| \ll m \cong E$). The four-component state vectors become
\begin{align}
u(p,1/2) &\;\cong\;
	\begin{pmatrix}
		1 \\ 0 \\ 0 \\ 0
	\end{pmatrix}\;, \quad
	\quad
u(p,-1/2) \;\cong\;
	\begin{pmatrix}
		0 \\ 1 \\ 0 \\ 0
	\end{pmatrix}
\label{eq:spinors_non_relativistic1}
\end{align}
and
\begin{align}
v(p,1/2) 	&\;\cong\;
	\begin{pmatrix}
		0 \\ 0 \\ 0 \\ 1 
	\end{pmatrix}\;, \quad
	\quad
v(p,-1/2) 	\;\cong\;
	\begin{pmatrix}
		0 \\ 0 \\ -1 \\ 0
	\end{pmatrix}\,.
\label{eq:spinors_non_relativistic2}
\end{align}

In this approximation, $T_{fi}$ can be written as
\begin{equation}
 T_{fi} = \frac{1}{(2 \pi)^6}\left(t_\text{exch}+t_\text{annihil}\right)\,.
\end{equation}
To compute the exchange term, we adopt the Dirac representation for the gamma matrices and the 
center-of-momentum frame, obtaining
\begin{align}
t_{\text{exch}} \,&=\, e^2\,\delta^{\mu 0} \delta_{\sigma_1 \tau_1} \,P_{\mu \nu}(k) \,\delta^{\nu 0} \delta_{\sigma_2 \tau_2} \nonumber \\ 
		\,&=\, e^2\, P_{0 0}(k)\, \delta_{\sigma_1 \tau_1} \delta_{\sigma_2 \tau_2}\,,
\label{eq:texch}
\end{align}
with\footnote{We are using the metric $g_{\mu \nu} = \diag(-1,1,1,1)$, which will be 
more convenient when we consider Wick rotations later.}
\begin{equation}
k \,=\, p_1\, - \,q_1\, = \begin{pmatrix} 0 ,& \vec{k}\end{pmatrix}\,.
\label{eq:transferred_momentum}
\end{equation}
For the annihilation term, note that conservation of momentum at the vertices implies that
\begin{equation}
k \,=\, \begin{pmatrix} 2 m ,& 0\end{pmatrix}\,.
\label{eq:transferred_momentum_annih}
\end{equation}
For QED, the Feynman-gauge propagator is given by the expression $P_{\mu \nu}(k) = - g_{\mu \nu}/k^2$. 
As seen in Eqs.\ (\ref{eq:spinors_non_relativistic1}) and (\ref{eq:spinors_non_relativistic2}), the 
spinors are momentum-independent
in the nonrelativistic approximation and, while $t_{\rm exch}$ is proportional to
$1/\vec{k}^2$, we see that $t_{\rm annihil}$ will be proportional to $1/4 m^2$.
Thus, we can neglect annihilation effects and the scattering amplitude is given by
\begin{equation}
T_{fi} = \frac{1}{(2 \pi)^6}\frac{e^2}{\vec{k}^2}\,.
\label{eq:scatter_amp}
\end{equation}
 
The potential can then be obtained as an inverse Fourier transform, which leads to the 
Coulomb potential
\begin{equation}
\begin{aligned}
 V(\vec{r}) &	\,=\, - (2 \pi)^3 \int \exp(-i \vec{k}\cdot\vec{r})\;T_{fi}(\vec{k}^2)\,d^3 k \\ & \,=\,- \frac{e^2}{4 \pi 	r}\,.
\end{aligned}
\label{eq:Couloumb_Pot}
\end{equation}

\vskip 3mm
For QCD, we replace the photon by the gluon and the electron-positron pair by a quark-antiquark pair.
The scattering amplitude will continue to be expressed as a sum of the two terms, now given by
\begin{align}
 \label{eq:QCD_Scattering_Amplitude}
t_{\text{exch}} \,&=\,\phantom{\,+\,}g_0^2\;\overline{u}(q_1,\tau_1)\,c_{1,\,f}^\dagger\,\lambda^a\gamma^\mu \,c_{1,\,i}\,u(p_1,\sigma_1)\;P_{\mu \nu}^{a b}(k)\; \nonumber \\[1mm]
		              &\phantom{ = \,+\,}\times\, \overline{v}(p_2,\sigma_2)\,c_{2,\,i}^\dagger\,\lambda^b\gamma^\nu\,c_{2,\,f}\,v(q_2,\tau_2)
\end{align}
and
\begin{align}
t_{\text{annihil}} \,&=\,-\,g_0^2\;\overline{v}(p_2,\sigma_2)\,c_{2,\,f}^\dagger\,\lambda^a\gamma^\mu  \,c_{1,\,i}\,u(p_1,\sigma_1)\;P_{\mu \nu}^{a b}(k)\; \nonumber \\[1mm]
	      &\phantom{=\,-\,}\times\,\overline{u}(q_1,\tau_1)\,c_{1,\,f}^\dagger\,\lambda^b\gamma^\nu c_{2,\,f}\,v(q_2,\tau_2)\,,
\end{align}
where $c_{(1,2),(i,f)}$ are three-component color vectors and $\lambda^a$ are
the Gell-Mann matrices. 

Let us note that, with respect to the QED case, the terms $t_{\rm exch}$ and $t_{\text{annihil}}$,
which contribute to the scattering amplitude $T_{fi}$ in Eq.\ (\ref{eq:QED_Scattering_Amplitude}),
now have multiplicative (Casimir) factors, coming from the sum over colors. This sum is obtained assuming that the 
incoming/outgoing quarks and antiquarks have equal probability of being in a given color state and
imposing a color-diagonal gluon propagator. 
Then, these factors are given respectively by
\begin{align}
\label{eq:color_factor_exch}
 c^\dagger_{1,f}\,\lambda^a\,c_{1,i}\;c^\dagger_{2,i}\,\lambda^a\,c_{2,f} = 
 \frac{1}{3} \Tr \lambda^a \lambda^a = \frac{\delta^{a a}}{6} = \frac{4}{3}\,
\end{align}
and
\begin{align}
\label{eq:color_factor_annih}
c^\dagger_{2,i}\,\lambda^a\,c_{1,i}\;c^\dagger_{1,f}\,\lambda^a\,c_{2,f} =
 \frac{1}{3} (\Tr \lambda^a) (\Tr \lambda^a) = 0\,.
\end{align}

Therefore, annihilation effects do not contribute, independently of the nonrelativistic approximation.
If we now assume a free (i.e.\ tree-level) gluon propagator
\begin{equation}
P_{\mu \nu}^{a b} = -\frac{g_{\mu \nu}\, \delta^{a b}}{k^2}\,, 
\end{equation}
we obtain a Coulomb-like interquark potential
\begin{equation}
V(r) = -\frac{4}{3}\frac{g_0^2}{4 \pi r} = -\frac{4}{3}\frac{\alpha_s}{r}\,.
\label{eq:QCD_pert_pot}
\end{equation}

Notice that the above potential is non-confining.
This could have been expected, since we have performed a purely perturbative calculation, while confinement 
is a nonperturbative phenomenon. The confinement property can then be obtained by addition of a linear term,
as described in Section \ref{sec:Introduction},
leading to the Cornell, or Coulomb-plus-linear, potential \cite{Eichten:1978tg,Eichten:1979ms,Eichten:2002qv}
\begin{equation}
V(r) = -\frac{4}{3}\frac{\alpha_s}{r} \,+\, \sigma\,r\,,
\label{eq:cornell_pot}
\end{equation}
which describes surprisingly well the states of charmonium and bottomonium.

\vskip 3mm
As mentioned in Section \ref{sec:Introduction}, we substitute the free propagator by a 
fully nonperturbative one in the OGE term. More precisely, we use the 
propagator\footnote{The energy scale used to convert $s$, $t$, and $u$ 
from lattice to physical units was set using
the value $\sqrt{\sigma}=0.44$ GeV for the string tension.
\protect\label{footsigma}}
\begin{align}
\label{eq:lattice_gluon_propagator}
 P_{\mu \nu}^{a b}(k) = \frac{C\,(s+k^2)}{t^2+u^2 k^2+k^4} \left(\delta_{\mu \nu} - \frac{k_\mu k_\nu}{k^2}\right) \delta^{a b}\,, \\[2mm]
  \begin{aligned}
C & = \num{0.784},\; & s &  = \SI{2.508}{\giga \electronvolt \squared}\,, \nonumber \\
t & = \SI{0.720}{\giga \electronvolt \squared}
,\; 
& u & = \SI{0.768}{\giga \electronvolt}
\,, \end{aligned}
\end{align}
obtained from fits of 
lattice data for a pure $SU(2)$ gauge theory in Landau gauge 
\cite{Cucchieri:2011ig}.
Note that the above parameters correspond to a
value $1/k^2$ at 2 \si{\giga \electronvolt}. Here we choose to normalize the propagator to $1/k^2$ at $k\to\infty$, i.e.\ we
adopt $C=1$.

We now follow the same procedure as in the QED case. From Eq.\ (\ref{eq:texch}) we notice that, in
the nonrelativistic approximation, only the component $P_{0 0}(0,\vec{k})$ survives in the $t_{\text{exch}}$ term and thus
the term $k_\mu k_\nu/k^2$ vanishes [see Eq.\ (\ref{eq:transferred_momentum})]. Lastly, in order to convert the propagator in 
Eq.\ (\ref{eq:lattice_gluon_propagator}), which was 
evaluated in Euclidean space,
to Minkowski space, we undo the Wick rotation, taking $\delta_{\mu \nu} \rightarrow -g_{\mu \nu}$. We obtain\footnote{Let 
us recall that the propagator is a gauge-dependent quantity. A gauge-independent
potential obtained from the (Coulomb-gauge) propagator is discussed in \cite{Popovici:2010fy}.}
\begin{equation}
 P_{0 0}^{a b}\big(\vec{k}\,\big) = \frac{C\,\big(s+\vec{k}^2\big)}{t^2+u^2 \vec{k}^2+\vec{k}^4}\, 
 \delta^{a b}\,.
\label{eq:Nonrelativistic_Lattice_Propagator}
\end{equation}
This leads us to the following scattering amplitude
\begin{equation}
T_{fi} = \frac{4}{3}\,\frac{g_0^2}{(2 \pi)^6}\,\frac{C\,\big(s+\vec{k}^2\big)}{t^2+u^2 \vec{k}^2+\vec{k}^4}\,.
\label{eq:scatter_amp_QCD}
\end{equation}

The potential is obtained, as was done in the QED case
[see Eq.\ (\ref{eq:Couloumb_Pot})], as a Fourier transform of the 
scattering amplitude $T_{fi}$.
We use spherical coordinates for $\vec{k}$ and set $\vec{r} = r\, \hat{z}$.
The angular integration is then trivial, resulting in\footnote{For the evaluation of this integral only, we
will denote $\abs{\vec{k}}=k$.}
\begin{multline}
V(\vec{r}\,) \,=\, -\frac{4}{3}\,\frac{\alpha_s}{r}\,\frac{C}{2\pi i}
\\ \times 
\int_{-\infty}^\infty{\frac{\left(s+k^2\right) 
\left(e^{i k r}-e^{-i k r}\right)}{t^2+u^2 k^2+k^4} k\,dk}\,,
\label{eq:angular_integration}
\end{multline}
where $\alpha_s = g_0^2/4 \pi$ [see Eq.\ (\ref{eq:QCD_pert_pot})].
The integral in Eq.\ (\ref{eq:angular_integration}) can be solved using residue calculations. The four poles 
in the integrand are symmetrically distributed in the four quadrants of the complex
plane. We index these poles in the following way
\begin{align}
k_{m,n} \;&=\; (-1)^m\, i \sqrt{t}\, \exp{\left[(-1)^n\,i \frac{\theta}{2}\right]}\,,\;\, m, \,n = 0,1\,,
\end{align}
where
\begin{align}
 \theta \;&\equiv\; \arctan \left(\frac{\sqrt{4 t^2-u^4}}{u^2}\right)\,.
\end{align}
The associated contour integral is performed by considering its two terms separately:
for the term with $e^{i k r}$ (respectively with $e^{-i k r}$) we close the contour 
above (respectively below). The residues are given by
\begin{multline}
 \Res \left[\frac{\left(s+k^2\right)k\,e^{\pm i k r}}{t^2+u^2 k^2+k^4},\, k_{m,n}\right] 
 = \\ \frac{1}{2}\,\frac{\left(s+k_{m,n}^2\right)\,e^{\pm i k_{m,n} r}}{ u^2+2 k_{m,n}^2}\,.
\end{multline}
The result is simplified by noticing that 
$\,k_{1,0} = -k_{0,0}$, $\,k_{0,1}=-k_{1,1}$ and $\,k_{1,1}=k_{0,0}^*$. 
The obtained potential, which we call the lattice-gluon-propagator 
potential $V_{LGP}$, is then
\begin{equation}
V_{LGP}(r) = -\frac{4}{3} \frac{\alpha_s}{r}
 \Re \left[ \frac{2 C(s+k_{0,0}^2)\,e^{i k_{0,0} r}}{u^2+2 k_{0,0}^2}\right]\,,
\label{eq:Potential_QCD}
\end{equation}
where $\Re$ indicates the real part. 

\begin{figure}[t]
\centering
\caption{Comparison between the lattice-gluon-propagator potential $V_{\mbox{LGP}}$ and the Coulomb-like 
potential (color factor included) in the charmonium case.}
\label{fig:Comparison_Potentials}
\vspace{2mm}
 \includegraphics{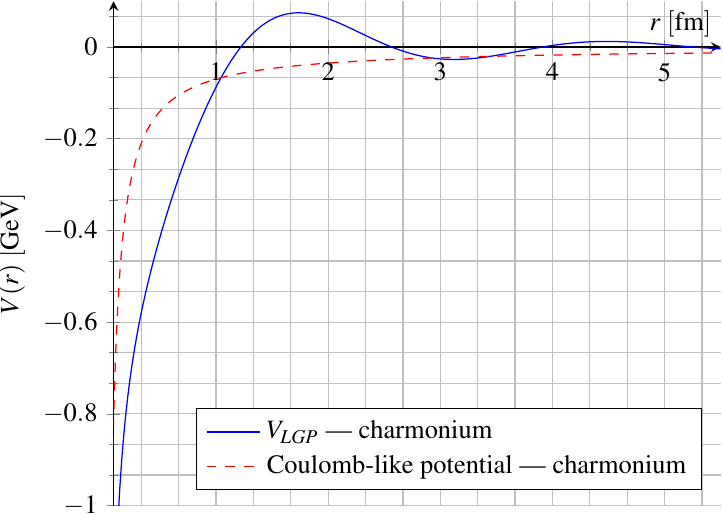}
\end{figure}

We note that the only difference with respect to the perturbative 
(Coulomb-like) case [see Eq.\ (\ref{eq:QCD_pert_pot})] is given by the 
expression within brackets.
In order to get a quantitative comparison between the two results, 
we only need to set the value of the strong coupling constant
$\alpha_s$ in Eqs.\ (\ref{eq:QCD_pert_pot}) and (\ref{eq:Potential_QCD}).
To this end, we evaluate $\alpha_s$ ---at the energy scale of the mass of 
the 1S quarkonia states [respectively $J/\psi$ and $\Upsilon$(1S) in the 
charmonium and bottomonium cases]--- by using the four-loop formula and the 
$\Lambda_{\mbox{QCD}}$ values in Ref.\ \cite[Section 9]{Agashe:2014kda}.
This yields $\alpha_s \approx 0.2663$ for the charmonium and
$\alpha_s \approx 0.1843$ for the bottomonium.
The resulting potentials are compared (for the charmonium case) 
in Fig.\ \ref{fig:Comparison_Potentials}.

We see that the two curves are clearly different, with $V_{LGP}(r)$ rising above zero at around the
hadronic scale (i.e.\ for $r\approx 1$ fm). For larger distances, the curve drops and it
can be observed that the potential $V_{LGP}$ is also non-confining. Thus, since (tree-level) 
perturbation theory was applied, the property of confinement was lost, even though the used 
propagator was obtained nonperturbatively. Nevertheless, one may hope to describe the first few
bound states of the spectrum solely using $V_{LGP}$.
This is done in Section \ref{sec:Results}.
We also consider the addition of a linearly rising term 
$\sigma \, r$ to the potential in order to model confinement, as done for the Cornell-potential case. 
In this case, the resulting expression is the 
lattice-gluon-propagator-plus-linear potential
\begin{equation}
V_{LGP+L}(r)\;\equiv\;V_{LGP}(r)\,+\,\sigma\,r\,.
\label{eq:lgp+l_potential}
\end{equation}
Note that the nonrelativistic approximation removes any spin dependence from the interactions. 
This means that, in our description, states with different spin values will be degenerate.

\section{Numerical Method}
\label{sec:Methods}
 
Let us consider a central (nonrelativistic) potential describing the 
interaction between two particles. 
Since we are dealing with a two-particle system, we can write 
the Hamiltonian in terms of relative coordinates and use separation of variables 
in the resulting partial differential equation to isolate the angular part 
of the wave function, given by the spherical harmonics. Lastly, 
we perform the usual substitution of variables in the radial wave function
$R(r) = f(r)/r$ to obtain the ordinary differential equation (ODE) for $f(r)$
\begin{equation}
\label{eq:Radial_Equation}
\frac{d^2 f}{dr^2}\,+\,2\mu\left[E - V(r) -2m - \frac {l\left(l+1\right)}{2 \mu r^2} \right] f\left(r\right) = 0 \,,
\end{equation}
where $\mu$ is the reduced mass
\begin{equation}
\mu = \frac{m}{2}\,,
\end{equation}
$l$ is the quantum number associated with the angular momentum
and $m$ is the mass of the heavy (charm or bottom) quark.
We use units such that $c = \hbar = 1$.
Notice as well the addition of the rest mass of the particles, which will allow us to 
compare the eigenvalue directly with the mass values given in Ref.\ \cite{Agashe:2014kda}.

The above ODE has to be solved with proper boundary-value conditions. The first condition is that
$f(0) = 0$. This comes from the requirement that $R(0)$ be non-singular. A second
condition is that $f(r \to \infty) = 0$ and comes from the fact that $R(r)$ is
normalized, i.e.
\begin{equation}
\int_0^\infty \left| R(r) \right|^2 r^2 dr = \int_0^\infty \left| f(r) \right|^2 dr = 1\,.
\end{equation}

In the limit of large $r$, the potential is
dominated by the linearly rising term and the ODE becomes
\begin{equation}
	\frac{d^2 f(r)}{dr^2}\,-\,2\mu \sigma r f\left(r\right) = 0 \,.
	\label{eq:ODE_large_r}
\end{equation}

The general solution of this equation is the linear combination of the Airy functions $Ai(\rho)$ and 
$Bi(\rho)$ \cite{abramowitz1964handbook}, where
$\,\rho = (2 \mu \sigma)^{1/3} \,r$. However, the Airy function of the second kind $\Bi(\rho)$ diverges at 
large $\rho$ and therefore it does not obey the boundary condition at infinity. 
For $\rho>0$, the Airy function of the first kind can be written as
\begin{equation}
	\Ai(\rho) = \frac{1}{\pi}\sqrt{\frac{\rho}{3}} \BesselK_{1/3}\left(\frac{2}{3}\rho^{3/2}\right)\,,
	\label{eq:Airy_Function}
\end{equation}
where $\BesselK_\nu(x)$ is the modified Bessel function of the second kind. One can try the Ansatz 
$f(\rho) = g(\rho) \Ai(\rho)$ and use the property $\BesselK_\nu'(x) = \nu \BesselK_\nu(x)/x-K_{\nu+1}$
in the ODE in Eq.\ (\ref{eq:Radial_Equation}) to obtain a second-order ODE with coefficients in terms of 
$\Ai(\rho)$ and $\BesselK_{4/3}(x)$. However, by expressing these functions as a power series in $\rho$,
one clearly sees that an analytic solution would be challenging, even for 
the simpler case of the Cornell potential. 
We therefore seek a numerical solution of the problem.

\begin{table}
\caption{Ranges of parameter values and iterative step sizes
used to obtain the charmonium and bottomonium eigenenergies $E$ with
the $V_{LGP}$ potential.
We also show the range and step of integration for the radial
distance $r$.}
\label{tb:Input_Parameters0}
\begin{tabular}{cccc}
\hline \hline
						& Charmonium 																	 & Bottomonium																	 & Step \\ \hline
E & \hskip 3mm \num{2.50} to \SI{4.50}{\giga \electronvolt} &\hskip 3mm  \num{8.50} to \SI{12.50}{\giga \electronvolt}  & \SI{0.04}{\giga \electronvolt} \\
$m_c$				& \num{1.00} to \SI{2.25}{\giga \electronvolt} &   																						 & \SI{0.001}{\giga \electronvolt} \\
$m_b$				&	  																					 & \num{4.00} to \SI{6.00}{\giga \electronvolt}  & \SI{0.001}{\giga \electronvolt} \\
$\alpha_s$	& \multicolumn{2}{c}{ \num{0.10} to \num{1.00} } & \num{0.01}  \\
$r$					& \multicolumn{2}{c}{\num{200.0} to \SI{0.0}{\per \giga \electronvolt} } & \SI{0.05}{\per \giga \electronvolt} \\ \hline \hline
\end{tabular}
\end{table}

\begin{table}
\caption{Ranges of parameter values and iterative step sizes
used to obtain the charmonium and bottomonium eigenenergies $E$ with
the $V_{LGP+L}$ and Cornell potentials. For $\alpha_s$, taken as a 
fixed parameter, we list the corresponding perturbative values (see 
Section \ref{sec:Pot_Model_Review}).
We also show the range and step of integration for the radial
distance $r$.}
\label{tb:Input_Parameters}
\begin{tabular}{cccc}
\hline \hline
						& Charmonium 																	 & Bottomonium																	 & Step \\ \hline
E & \hskip 3mm \num{2.00} to \SI{6.00}{\giga \electronvolt} &\hskip 3mm  \num{8.50} to \SI{12.50}{\giga \electronvolt} & \SI{0.04}{\giga \electronvolt} \\
$m_c$				& \num{1.00} to \SI{2.00}{\giga \electronvolt} &   																						 & \SI{0.01}{\giga \electronvolt} \\
$m_b$				&	  																					 & \num{4.15} to \SI{4.85}{\giga \electronvolt}  & \SI{0.01}{\giga \electronvolt} \\
$\sigma$		& \multicolumn{2}{c}{\num{0.10} to \SI{0.50}{\giga \electronvolt \squared}} & \SI{0.01}{\giga \electronvolt \squared} \\
$\alpha_s$	& \num{0.2663} & \num{0.1843} &   \\
$r$					& \multicolumn{2}{c}{\num{50.0} to \SI{0.0}{\per \giga \electronvolt} } & \SI{0.0125}{\per \giga \electronvolt} \\ \hline \hline
\end{tabular}
\end{table}

\begin{table*}[t]
\centering
\caption{Experimental spectrum of charmonium states and spin-averaged values
using two different methods (see text).}
\label{tb:input_data_charm}
\begin{tabular}{cS[table-format=2.9]cccc}
\hline
\hline
Particle Name 		& {Mass (\si{\giga \electronvolt})}	& $J^{P C}$	& $l$ 			& {MAV1 (\si{\giga \electronvolt})}				& {MAV2 (\si{\giga \electronvolt})}			\\ \hline
$\eta_c(1S)$		& \EtacOneS +- \EtacOneSErr		& $0^{- +}$	& \multirow{2}{*}{0}	& \multirow{2}{*}{\tablenum{\InpCharmOneS +- \InpCharmOneSErr}}	& \multirow{2}{*}{\tablenum{3.040 +- .057}}\\
$J/\psi(1S)$		& \PsiOneS +- \PsiOneSErr		& $1^{- -}$	& 			& 									& \\ \hline
$\chi_{c0}(1P)$ 	& \chicZOneP +- \chicZOnePErr 		& $0^{+ +}$	& \multirow{4}{*}{1}	& \multirow{4}{*}{\tablenum{\InpCharmOneP +- \InpCharmOnePErr}}	& \multirow{4}{*}{\tablenum{3.485  +- .070}}\\
$\chi_{c1}(1P)$		& \chicOOneP +- \chicOOnePErr		& $1^{+ +}$	& 			&									& \\
$h_c(1P)$		& \hcOneP +- \hcOnePErr			& $1^{+ -}$	& 			&									& \\
$\chi_{c2}(1P)$		& \chicTOneP +- \chicTOnePErr		& $2^{+ +}$	& 			&									& \\ \hline
$\eta_c(2S)$		& \EtacTwoS +- \EtacTwoSErr		& $0^{- +}$	& \multirow{2}{*}{0}	& \multirow{2}{*}{\tablenum{\InpCharmTwoS +- \InpCharmTwoSErr}}	& \multirow{2}{*}{\tablenum{ 3.663 +- .023}}\\
$\psi(2S)$		& \PsiTwoS +- \PsiTwoSErr		& $1^{- -}$	& 			& 									& \\ \hline
$\psi(3770)$		& \PsiA +- \PsiAErr			& $1^{- -}$	&	0 or 2		& \multirow{1}{*}{\tablenum{\PsiA +- \PsiAErr}}			& \multirow{1}{*}{\tablenum{\PsiA +- \PsiAErr}} \\ \hline
$X(3872)$		& \XA +- \XAErr				& $1^{+ +}$	& 1 			& \multirow{1}{*}{\tablenum{\XA +- \XAErr}}				& \multirow{1}{*}{\tablenum{\XA +- \XAErr}} \\ \hline
$\chi_{c0}(2P)$		& \chicZTwoP +- \chicZTwoPErr		& $0^{+ +}$	& \multirow{2}{*}{1}	& \multirow{2}{*}{\tablenum{3.9257 +- 0.0025}}				& \multirow{2}{*}{\tablenum{ 3.9228 +-0.0044}}\\ 
$\chi_{c2}(2P)$		& \chicTTwoP +- \chicTTwoPErr		& $2^{+ +}$	& 			& 									& \\ \hline
$\psi(4040)$		& \PsiB +- \PsiBErr			& $1^{- -}$	& 0 or 2		& \multirow{1}{*}{\tablenum{\PsiB +- \PsiBErr}}			& \multirow{1}{*}{\tablenum{\PsiB +- \PsiBErr}}\\ \hline
$\psi(4160)$		& \PsiC +- \PsiCErr			& $1^{- -}$	& 0 or 2		& \multirow{1}{*}{\tablenum{\PsiC +- \PsiCErr}}			& \multirow{1}{*}{\tablenum{\PsiC +- \PsiCErr}}	\\ \hline
$X(4260)$ 		& \XB +- \XBErr				& $1^{- -}$	& 0 or 2		& \multirow{1}{*}{\tablenum{\XB +- \XBErr}}				& \multirow{1}{*}{\tablenum{\XB	+- \XBErr}}\\ \hline
$X(4360)$ 		& \XC +- \XCErr				& $1^{- -}$	& 0 or 2		& \multirow{1}{*}{\tablenum{\XC +- \XCErr}}				& \multirow{1}{*}{\tablenum{\XC +- \XCErr}}\\ \hline
$\psi(4415)$		& \PsiD +- \PsiDErr			& $1^{- -}$	& 0 or 2		& \multirow{1}{*}{\tablenum{\PsiD +- \PsiDErr}}			& \multirow{1}{*}{\tablenum{\PsiD +- \PsiDErr}}\\ \hline
$X(4660)$		& \XD +- \XDErr				& $1^{- -}$	& 0 or 2		& \multirow{1}{*}{\tablenum{\XD +- \XDErr}}				& \multirow{1}{*}{\tablenum{\XD	+- \XDErr}}\\ \hline \hline
\end{tabular}
\end{table*}
\begin{table*}[t]
\centering
\caption{Experimental spectrum of bottomonium states and spin-averaged values
using two different methods (see text).
We include the unconfirmed state $\eta_b(1S)$.}
\label{tb:input_data_bottom}
\begin{tabular}{cS[table-format=2.9]cccc}
\hline \hline
Particle Name 		& {Mass (\si{\giga \electronvolt})}	& $J^{P C}$	& $l$ 			& {MAV1 (\si{\giga \electronvolt})}				& {MAV2 (\si{\giga \electronvolt})}			\\ \hline
$\eta_b(1S)$		& \EtabOneS +- \EtabOneSErr		& $0^{- +}$	& \multirow{2}{*}{0}	& \multirow{2}{*}{\tablenum{9.4447 +- 0.0010}}				& \multirow{2}{*}{\tablenum{9.429 +- .031}}\\ 
$\Upsilon(1S)$		& \UpsilonOneS +- \UpsilonOneSErr	& $1^{- -}$	& 			&									& \\ \hline
$\chi_{b0}(1P)$		& \chibZOneP +- \chibZOnePErr		& $0^{+ +}$	& \multirow{4}{*}{1}	& \multirow{4}{*}{\tablenum{\InpBottomOneP +- \InpBottomOnePErr}}	& \multirow{4}{*}{\tablenum{ 9.886 +- .026}}\\
$\chi_{b1}(1P)$		& \chibOOneP +- \chibOOnePErr		& $1^{+ +}$	&			&									& \\
$h_b(1P)$		& \hbOneP +- \hbOnePErr			& $1^{+ -}$	&			&									& \\
$\chi_{b2}(1P)$		& \chibTOneP +- \chibTOnePErr  	& $2^{+ +}$	&			&									& \\ \hline
$\Upsilon(2S)$		& \UpsilonTwoS +- \UpsilonTwoSErr	& $1^{- -}$	&	0		& \multirow{1}{*}{\tablenum{\UpsilonTwoS +- \UpsilonTwoSErr}}		& \multirow{1}{*}{\tablenum{\UpsilonTwoS +- \UpsilonTwoSErr}}\\ \hline
$\Upsilon(1D)$		& \UpsilonOneD +- \UpsilonOneDErr	& $2^{- -}$	&	2		& \multirow{1}{*}{\tablenum{\UpsilonOneD +- \UpsilonOneDErr}}		& \multirow{1}{*}{\tablenum{\UpsilonOneD +- \UpsilonOneDErr}}\\ \hline
$\chi_{b0}(2P)$		& \chibZTwoP +- \chibZTwoPErr		& $0^{+ +}$	& \multirow{3}{*}{1}	& \multirow{3}{*}{\tablenum{\InpBottomTwoP +- \InpBottomTwoPErr}}	& \multirow{3}{*}{\tablenum{ 10.251 +- .018}}\\
$\chi_{b1}(2P)$		& \chibOTwoP +- \chibOTwoPErr		& $1^{+ +}$	&			& 									& \\
$\chi_{b2}(2P)$		& \chibTTwoP +- \chibTTwoPErr		& $2^{+ +}$	&			& 									& \\ \hline
$\Upsilon(3S)$		& \UpsilonThreeS +- \UpsilonThreeSErr	& $1^{- -}$	&	0		& \multirow{1}{*}{\tablenum{\UpsilonThreeS +- \UpsilonThreeSErr}}	& \multirow{1}{*}{\tablenum{\UpsilonThreeS +- \UpsilonThreeSErr}}\\ \hline
$\chi_{b}(3P)$		& \chibThreeP +- \chibThreePErr	& $?^{? +}$	& 1			& \multirow{1}{*}{\tablenum{\chibThreeP +- \chibThreePErr}}		& \multirow{1}{*}{\tablenum{\chibThreeP +- \chibThreePErr}}\\ \hline	
$\Upsilon(4S)$		& \UpsilonFourS +- \UpsilonFourSErr	& $1^{- -}$	&	0		& \multirow{1}{*}{\tablenum{\UpsilonFourS +- \UpsilonFourSErr}}	& \multirow{1}{*}{\tablenum{\UpsilonFourS +- \UpsilonFourSErr}}\\ \hline
$\Upsilon(10860)$	& \UpsilonA +- \UpsilonAErr		& $1^{- -}$	& 0 or 2		& \multirow{1}{*}{\tablenum{\UpsilonA +- \UpsilonAErr}}		& \multirow{1}{*}{\tablenum{\UpsilonA +- \UpsilonAErr}}\\ \hline
$\Upsilon(11020)$	& \UpsilonB +- \UpsilonBErr		& $1^{- -}$	& 0 or 2		& \multirow{1}{*}{\tablenum{\UpsilonB +- \UpsilonBErr}}		& \multirow{1}{*}{\tablenum{\UpsilonB +- \UpsilonBErr}}\\ \hline \hline
\end{tabular}
\end{table*}

To this end, we use the so-called shooting method \cite{press2007numerical}. It consists in picking trial values in a discretized
range for the eigenenergies, integrating the ODE for each of these values to 
obtain the corresponding wave function, and choosing the energies for which 
the wave function obeys the boundary conditions approximately. 
We use the backward second-order Runge-Kutta method to integrate the wave
function, starting from a maximum value $r_{max}$ for the radial coordinate until the origin, in steps 
of $dr$ (we adapt the method as presented in Ref.\ \cite{press2007numerical} 
by adopting a negative integration step). We choose $r_{max}$ sufficiently 
large so that we can use $f(r_{max}) = \Ai(r_{max})$ and $f'(r_{max}) = \Ai'(r_{max})$ as initial conditions. 
In practice, the wave 
function will not obey the boundary conditions exactly since the proposed energy is unlikely to be an exact eigenenergy. 
Nevertheless, we may count the number of nodes of the wave function: each time we observe an increase in the 
number of nodes when compared with the previously proposed energy, the desired eigenenergy will be between the 
two proposed values. 
We further refine our method by adapting the bisection method to search for the eigenenergy in this interval, 
thus allowing the use of a coarse grid without loss of precision.

We first test the above method for the Cornell potential [see Eq.\ 
(\ref{eq:cornell_pot})], with 
parameters fixed to 
$\sigma = \SI{1}{\giga \electronvolt \squared}$, $2\mu = \SI{1}{\giga \electronvolt} $ and $4 \alpha_s/3 = 1$. 
These are the values used in Ref.\ \cite{2014arXiv1411.2023H}, 
which adopts a different approach (the asymptotic iteration method) 
for solving the problem.
We find agreement with their values
up to the 4th and in some cases even 5th decimal place. Similarly, Ref.\ \cite{Chung:2008sm} uses yet another 
numerical method to compute the eigenenergies for a different set of 
parameters, allowing comparison with our results. In this case 
we find agreement up to the 3rd decimal places. We must consider that, 
in this comparison, our parameters are close to but not identical
to the ones used in Ref.\ \cite{Chung:2008sm},
which might explain the slightly worse agreement than in the comparison 
with Ref.\ \cite{2014arXiv1411.2023H}.

\subsection{Fitting Procedure}

In general, we consider an expression for the potential with free parameters, 
to be fitted to a few experimental values.
To find the best fit, we set up a grid of values for these parameters. 
Then, we compute the eigenenergies for each proposed 
set of parameters and select the one that best describes the observed spectrum.
As a criterion for choosing the optimal parameters we consider the 
minimization of $\chi^2$ in the description of a few input values from 
experiment, i.e.\ we pick the set of parameters minimizing
\begin{equation}
	\chi^2(\text{parameters}) = \sum_{i} \left(\frac{E_{i}-E_{i,\text{experimental}}}{\sigma_i}\right)^2\,,
	\label{eq:residual_criteria}
\end{equation}
where $\sigma_i$ is the experimental error associated with the energy 
$E_{i,\text{experimental}}$ and the $E_i$'s
are the eigenenergies computed numerically.
In order to establish a confidence level for our parameters, we use the 
method described in detail in Ref.\ \cite{press2007numerical}, which 
consists in determining the region in parameter space for which 
$\chi^2/d.o.f.$ increases by less than one unit with respect to its 
minimum value, for each of the parameters separately. 
In cases for which the obtained confidence level is asymmetric, we adopt the larger value as the error.

The above prescription indirectly allows us to establish confidence 
limits for the eigenenergies, by the
so-called Monte Carlo method \cite{press2007numerical}. 
More precisely, for each parameter,
we draw $N = 1000$ random numbers following a Gaussian distribution,
centered at the optimal value of the fitted parameter and with standard 
deviation given by the symmetrized error, and evaluate the spectrum for
each (generated) synthetic set of parameter values. The corresponding
set of eigenenergies is then used to estimate the confidence limits
for the bound-state masses.

\vskip 3mm
Notice that the procedure described here can be applied to any 
central potential. 
In the next section, we perform several calculations using this method,
considering two approximately nonrelativistic systems: 
charmonium and bottomonium.
Of course,
since bottomonium states are heavier in comparison with their kinetic 
energy, we expect to obtain better results in this case than for charmonium.

Let us now outline our choices for the interquark potentials, the fitting
parameters, and the experimental data used for input and comparison.
We start from the $V_{LGP}$ potential [see Eq.\ (\ref{eq:Potential_QCD})]
obtained purely from the lattice gluon propagator.
In this case, as free parameters in the fits, we take
the strong coupling constant $\alpha_s$ and the mass $m$ of the heavy quark. 
A motivation for including $m$ as a free parameter
is that quark masses are not observable directly and depend on the 
renormalization scheme.
The ranges of parameter values ($m$ and $\alpha_s$)
and corresponding step sizes used to find the eigenenergies in the case
of the $V_{LGP}$ potential are given in
Table \ref{tb:Input_Parameters0}. We also list the range and step 
of integration for the radial distance $r$.

Next, we consider the $V_{LGP+L}$ and Cornell potentials,
which share the same parameters [see Eqs.\ (\ref{eq:lgp+l_potential})
and (\ref{eq:cornell_pot})].
Here one has the string tension $\sigma$ as a possible additional 
parameter. 
In order to have a fair comparison, our
calculations are done with two free parameters for the three potentials
separately. Namely, for the $V_{LGP+L}$ and Cornell potentials,
we choose to leave $\sigma$ (which is of nonperturbative nature)
and $m$ free and to fix 
$\alpha_s$ to its perturbative value (at the appropriate
energy scale). Furthermore, we perform a combined (constrained)
fit of charmonium and bottomonium results, leaving as free parameters
the two heavy quark masses and $\sigma$. This is our preferred fit.
The ranges of parameter values (heavy-quark mass and string tension $\sigma$)
and corresponding step sizes used to find the eigenenergies in the case
of the $V_{LGP+L}$ and Cornell potentials, as well 
as the calculated (fixed) values of $\alpha_s$, are given in 
Table \ref{tb:Input_Parameters}. We also list the range and step 
of integration for the radial distance $r$.

\vskip 3mm
Regarding the choice of experimental data for bound-state masses, we recall
that spin interactions are not considered in our approach.
This implies energy values with high degeneracy (in comparison with the 
experimental data) and we thus average over states with different spin. 
A possible averaging
procedure, used in Ref.\ \cite{Olsson:1994cv}, is to take the degeneracy 
of each state as a weight. The spin-averaged mass of the states with 
principal quantum number $n$ and in the $X$-wave state ($X = S,P,D \dots$) 
is then given by 
\begin{equation}
\langle M(nX) \rangle \; = \; \frac{\sum_{i=1}^{N_l} m_i(nX) g_i}{\sum_{i=1}^{N_l} g_i}\,,
\label{eq:spin_average}
\end{equation}
where $m_i(nX)$ is the mass of each of the $N_l$ states with the same 
angular momentum $l$, and $g_i$ is the degeneracy of the state.
The uncertainty associated with the above average may be estimated by 
propagation of errors, taking the width of the resonance peak\footnote{We 
recall that bound states are identified by plotting a histogram of number 
of particles (cross-section) detected in a collision versus the energy of 
the collision. When a resonance is found, it is associated to a bound state.} 
as the uncertainty in each mass $m_i(nX)$.
We refer to this averaging procedure as ``MAV1''.
A second possibility to average over different spins is to imagine that,
if the experiments were not very precise, we would not see several narrow 
nondegenerate states, but broad degenerate ones, i.e.\ a low-precision 
experiment would see the peaks merged.
We thus take the spin-averaged mass from the midpoint between the state with 
lowest energy and the one with highest energy. The error is estimated as 
half of the distance between these two states.
We refer to this method as ``MAV2''. 

The results corresponding to the two averaging procedures described above are 
reported in Tables \ref{tb:input_data_charm} and \ref{tb:input_data_bottom} 
respectively for charmonium and bottomonium states.
As experimental data, we choose to include only the states present in 
the meson summary table of Ref.\ \cite{Agashe:2014kda} that are regarded as 
established particles. Also, we omit charged states from our tables, since 
quarkonia states must be neutral. As inputs in the fits, we use the 
states $1S$, $1P$ and $2S$ (of charmonium and bottomonium).

By fitting (independently) the charmonium and bottomonium spectra
using MAV1, for the $V_{LGP+L}$ and Cornell potentials,
we find very large values of $\chi^2/d.o.f.$,
varying from $1.9 \times 10^2/d.o.f.$ (for bottomonium and $V_{LGP+L}$)
to $23 \times 10^3/d.o.f.$ (for charmonium and Cornell potential). 
Instead, the procedure MAV2 gives acceptable
values for $\chi^2/d.o.f.$ in the charmonium case, and larger values for 
bottomonium (still, orders of magnitude smaller than with MAV1). 
These $\chi^2/d.o.f.$ values improve if an unconfirmed state of 
bottomonium is included [namely, the $\eta_b(1S)$].
We choose MAV2 as our preferred method.

\section{Results}
\label{sec:Results}

\begin{table}[t]
\caption{Results for the charmonium eigenstates using the $V_{LGP}$ potential.
We leave $\alpha_s$ and $m_c$ as free parameters, obtaining 
$\alpha_s = 0.95\pm 0.02$ and 
$m_c = \SI[parse-numbers = false]{2.064_{-0.009}^{+0.010}}{\giga \electronvolt}$.
The states 1S, 2S and 1P were used as inputs in the fits. 
Long dashes represent states that have not been observed experimentally.}
\label{tb:results_VLGP_charm}
\centering
\begin{tabular}{cS[table-format=2.5]S[table-format=2.5]S[table-format=2.5]S[table-format=2.5]}
\hline \hline
\multicolumn{3}{c}{Charmonium Spectrum} \\ \hline
\multicolumn{3}{c}{{$V_{LGP}$}}	\\ \hline
{State}	& {Mass}	& {Deviation from average}		\\
	    & {(\si{\giga \electronvolt})}	& {spin state (\si{\giga \electronvolt})}	\\ \hline
{1S} & 3.05 +-  .11 & 0.01 \\
{1P} & 3.64 +- .13 & -0.02 \\
{2S} & 3.64 +-  .13 & 0.16 \\
{1D} & 3.64	 +- .13 & $\!\!\!\!\mbox{---\phantom{oii}}$ \\
{2P} & 4.10 +- .13 & 0.18 \\
{3S} & 4.12 +-  .15 & $\!\!\!\!\mbox{---\phantom{oii}}$ \\
{2D} & 4.13	 +- .18 & $\!\!\!\!\mbox{---\phantom{oii}}$ \\
{3P} & 4.13 +- .24 & $\!\!\!\!\mbox{---\phantom{oii}}$ \\
{4S} & 4.13 +-  1.4 & $\!\!\!\!\mbox{---\phantom{oii}}$ \\ \hline \hline
\end{tabular}
\end{table}

\begin{table}[t]
\caption{Results for the bottomonium eigenstates using the $V_{LGP}$ potential.
We leave $\alpha_s$ and $m_b$ as free parameters, obtaining 
$\alpha_s = 0.513^{+0.009}_{-0.010}$ and 
$m_b = \SI[parse-numbers = false]{5.10947^{+0.00016}_{-0.00014}}{\giga \electronvolt}$.
The states 1S, 2S and 1P were used as inputs in the fits. 
Long dashes represent states that have not been observed experimentally.}
\label{tb:results_VLGP_bottom}
\centering
\begin{tabular}{cS[table-format=2.5]S[table-format=2.5]S[table-format=2.5]S[table-format=2.5]}
\hline \hline
\multicolumn{3}{c}{Bottomonium Spectrum} \\ \hline
\multicolumn{3}{c}{{$V_{LGP}$}}	\\ \hline
{State}	& {Mass}	& {Deviation from average}		\\
	    & {(\si{\giga \electronvolt})}	& {spin state (\si{\giga \electronvolt})}	\\ \hline
{1S}	& 9.43  +- .30		& 0.00\\
{1P}	& 10.02 +- .32		& 0.14\\
{2S}	& 10.02 +- .32		& 0.00\\
{1D}	& 10.02 +- .32		& -0.14\\
{2P}	& 10.13 +- .32		& -0.12\\
{3S}	& 10.16 +- .32		& -0.19\\
{3P}	& 10.22 +- .32		& -0.31\\
{4S}	& 10.22 +- .32		& -0.36\\ 
{2D}	& 10.22 +- .32		& $\!\!\!\!\mbox{---\phantom{oii}}$\\ \hline \hline
\end{tabular}
\end{table}

We now follow the procedure described in Section \ref{sec:Methods} and
obtain, for a given potential, higher eigenenergies of the spectrum
from fits to a few low-lying states.
A natural first attempt is to consider the $V_{LGP}$ potential 
in Eq.\ (\ref{eq:Potential_QCD}) for the charmonium and bottomonium spectra. 
As explained in the previous section,
we do this by leaving the strong coupling constant $\alpha_s$ and the
mass of the heavy quark as free parameters, taking
the 1S, 2S and 1P energy states as inputs in the fits.
Spin averages are done using the MAV2 method.
The corresponding results are presented in Tables 
\ref{tb:results_VLGP_charm} and \ref{tb:results_VLGP_bottom}. 
Note that we also show the difference between each evaluated mass and the
corresponding experimental value. 
Long dashes represent states that have not been observed experimentally.
As can be seen, although the potential is non-confining (see discussion
in Section \ref{sec:Pot_Model_Review}), the existence of a few lowest 
states is qualitatively reproduced in the spectrum.
This is in agreement with the study in \cite{Gonzalez:2012hx} 
for charmonium states using an equivalent approach. 

However, it is clear that the pure OGE potential $V_{LGP}$ is not
enough to model the spectrum beyond its lowest states, or even to 
provide a quantitative description of these states. In fact,
the spacings between energy levels are not compatible with
the experimental values, both for the higher (estimated) states and 
for the lower ones used as inputs. Moreover (see Tables 
\ref{tb:results_VLGP_charm} and \ref{tb:results_VLGP_bottom}), 
we find that the energy states ``saturate'' 
around a maximum value.
As for the fit parameters, we obtain
$\alpha_s = 0.95\pm 0.02$ and
$m_c = \SI[parse-numbers = false]{2.064_{-0.009}^{+0.010}}{\giga \electronvolt}$,
for the charmonium, and
$\alpha_s = 0.513^{+0.009}_{-0.010}$ and
$m_b = \SI[parse-numbers = false]{5.10947^{+0.00016}_{-0.00014}}{\giga \electronvolt}$,
for the bottomonium.
We note that, while the bottom quark mass $m_b$ is not very far from the
experimental one (see the third column in Table
\ref{tb:parameters_results_simultaneous_fit})
the charm quark mass $m_c$ is almost twice the experimental datum
(see again Table \ref{tb:parameters_results_simultaneous_fit}).
At the same time, in both cases, the value obtained for $\alpha_s$
is quite far from the perturbatively
estimated one (see Table \ref{tb:Input_Parameters}).

\vskip 3mm
In the remainder of this section, we thus use the potential $V_{LGP+L}$
[see Eq.\ (\ref{eq:lgp+l_potential})], obtained by the addition of a linearly
growing term to $V_{LGP}$, as well as the Cornell potential
[see Eq.\ (\ref{eq:cornell_pot})] to generate the spectra, 
and perform a comparison of the results with the experimental data.
The idea is to combine the feature of
an improved description of the short-distance behavior of the system, 
as found above using the $V_{LGP}$ potential, with the imposition of
a linear behavior at large distances, which should help in obtaining
the higher energy states.
We will carry out the spectrum calculation ---as explained at the end
of Section \ref{sec:Methods}--- using the same set of parameters
for the two potentials.
As above, we consider the MAV2 averaging method
(see Tables \ref{tb:input_data_charm} and \ref{tb:input_data_bottom}), 
including the unconfirmed $\eta_b(1S)$ state.

\begin{table}
\caption{Quark masses and string tension obtained from our preferred fit.
These parameters are used to obtain the spectrum in Tables 
\ref{tb:results_simultaneous_fit_charm} and 
\ref{tb:results_simultaneous_fit_bottom}.}
\label{tb:parameters_results_simultaneous_fit}
\centering
\resizebox{\columnwidth}{!}{
\begin{tabular}{ccc} \hline \hline
	$V_{LGP+L}$ & Cornell Potential & Quark Mass\\
	            &                   & in Ref.\ \cite{Agashe:2014kda} \\ \hline
	$m_c = \SI{1.16 +- 0.03}{\giga \electronvolt}$																& $m_c = \SI[parse-numbers = false]{1.11_{-0.02}^{+0.08}}{\giga \electronvolt}$ 		& $m_c = \SI{1.275 +- 0.025}{\giga \electronvolt}$\\[2mm]
	$m_b = \SI[parse-numbers = false]{4.61_{-0.01}^{+0.02}}{\giga \electronvolt}$	& $m_b = \SI[parse-numbers = false]{4.58_{-0.01}^{+0.04}}{\giga \electronvolt}$ 		& $m_b(\overline{MS}) = \SI{4.18 +- 0.03}{\giga \electronvolt}$\\[2mm]
	$\sigma = \SI{0.23 +- 0.01}{\giga \electronvolt}^2$															& $\sigma = \SI[parse-numbers = false]{0.26_{-0.03}^{+0.01}}{\giga \electronvolt}^2$	& $m_b(1S) = \SI{4.66 +- 0.03}{\giga \electronvolt}$\\[2mm]
	$\chi^2 = \num{6.20}$																													& $\chi^2 = \num{12.13}$																														&\\	\hline \hline
\end{tabular}
}
\end{table}

\begin{figure*}
	\centering
\caption{Experimental mass spectrum for charmonium and corresponding 
spin averages. We also show our results in the 
$V_{LGP+L}$ and Cornell-potential cases from the 
constrained fit, considering as input the 
states 1S, 2S and 1P of the spectra. Averages are taken using the MAV2
procedure.}
\label{fig:Mass_Spectrum_charm}
\includegraphics[width=\textwidth]{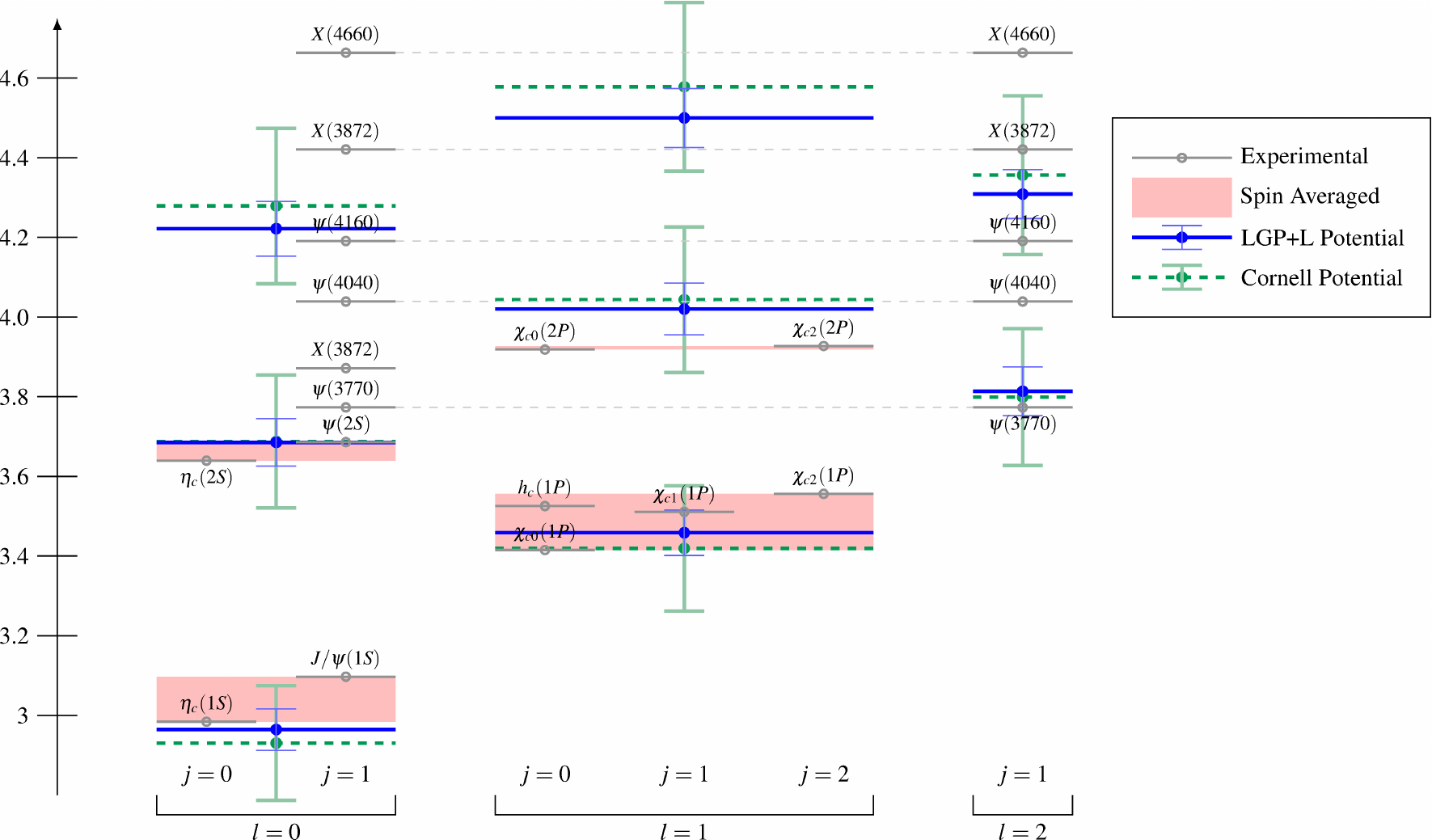}
\end{figure*}

\begin{figure*}
	\centering
\caption{Experimental mass spectrum for bottomonium and corresponding 
spin averages. We also show our results in the 
$V_{LGP+L}$ and Cornell-potential cases from the 
constrained fit, considering as input the 
states 1S, 2S and 1P of the spectra. Averages are taken using the MAV2
procedure.}
\label{fig:Mass_Spectrum_bottom}
\includegraphics[width=\textwidth]{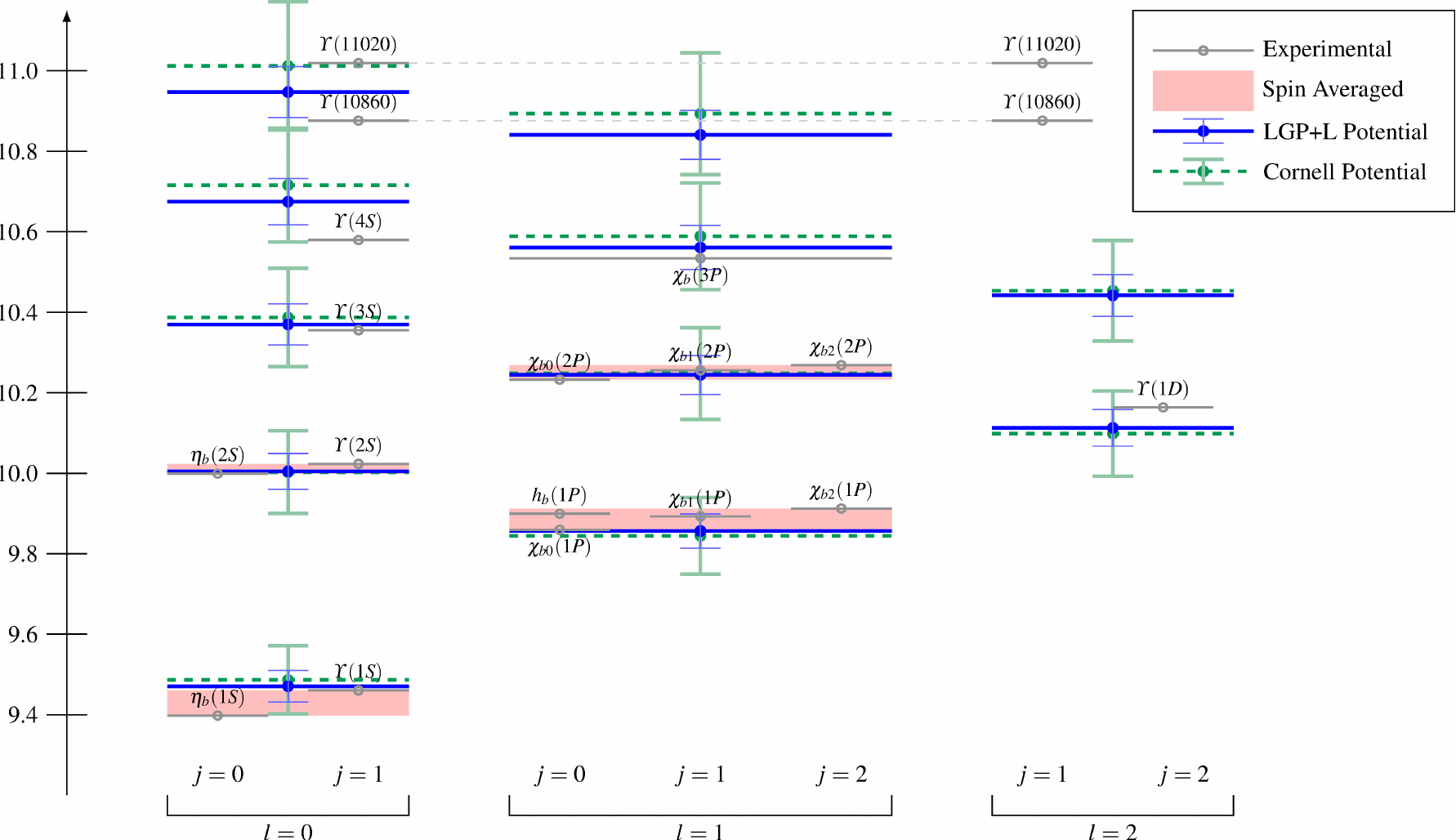}
\end{figure*}

The data obtained in the independent fits of charmonium and bottomonium spectra are then used to set up a 
constrained fit, i.e.\ one with a common value for the string tension $\sigma$ 
of the two systems.
Notice that, for this constrained fit, we have three
free parameters ($m_c$, $m_b$ and $\sigma$) and six states as inputs in the 
fit (the states 1S, 2S and 1P of charmonium and bottomonium), resulting in 
three degrees of freedom. The results of this fit using the $V_{LGP+L}$ 
and Cornell potentials are shown in Table
\ref{tb:parameters_results_simultaneous_fit}. 
The corresponding spectra are reported in Tables
\ref{tb:results_simultaneous_fit_charm} 
and \ref{tb:results_simultaneous_fit_bottom}. 
A visual representation of the spectra is provided in Figs.\ 
\ref{fig:Mass_Spectrum_charm} and \ref{fig:Mass_Spectrum_bottom}. 
Let us remark that the obtained value for the string tension $\sigma$
in our preferred fit (see Table \ref{tb:parameters_results_simultaneous_fit})
is rather close to the input value used to set the scale for the lattice gluon
propagator, $\sigma\approx 0.194$ GeV$^2$ (see Footnote \ref{footsigma}),
providing a nice consistency check.

For the charmonium spectrum, we obtain
smaller errors and (nevertheless) a slightly better agreement with the 
spin-averaged experimental values in the 
$V_{LGP+L}$ case than in the Cornell-potential one (see Table 
\ref{tb:results_simultaneous_fit_charm}).
In the bottomonium case the results obtained with the two
confining potential are comparable (see Table
\ref{tb:results_simultaneous_fit_bottom}).
Also, the central value for the string tension in the $V_{LGP+L}$
is slightly closer to the one used to set the energy scale for the
lattice propagator.

\begin{table}
\caption{Results for the charmonium eigenstates using $V_{LGP+L}$ and the 
Cornell potentials in a constrained fit (see text).
Long dashes represent states that have not been observed experimentally.}
\centering
\label{tb:results_simultaneous_fit_charm}
\begin{tabular}{cS[table-format=2.5]S[table-format=2.5]S[table-format=2.5]S[table-format=2.5]}
\hline \hline
\multicolumn{3}{c}{Charmonium Spectrum} \\ \hline
	\multicolumn{3}{c}{{$V_{LGP+L}$}} \\ \hline
{State}	& {Mass}	& {Deviation from average}		\\
	& {(\si{\giga \electronvolt})}	& {spin state (\si{\giga \electronvolt})}	\\ \hline
{1S}	& 2.96 +- .11		& -0.10\\
{1P}	& 3.46 +- .12		& -0.07\\
{2S}	& 3.69 +- .13		& 0.01\\
{1D}	& 3.81 +- .14		& $\!\!\!\!\mbox{---\phantom{oii}}$\\
{2P}	& 4.02 +- .14		& 0.09\\
{3S}	& 4.22 +- .15		& $\!\!\!\!\mbox{---\phantom{oii}}$\\
{2D}	& 4.31 +- .15		& $\!\!\!\!\mbox{---\phantom{oii}}$\\
{3P}	& 4.50 +- .16		& $\!\!\!\!\mbox{---\phantom{oii}}$\\
{4S}	& 4.69 +- .17		& $\!\!\!\!\mbox{---\phantom{oii}}$\\	\hline
\multicolumn{3}{c}{{Cornell Potential}} \\ \hline
{State}	& {Mass}	& {Deviation from average}		\\
	& {(\si{\giga \electronvolt})}	& {spin state (\si{\giga \electronvolt})}	\\ \hline
{1S}	& 2.93 +- .17		& -0.14\\
{1P}	& 3.42 +- .19		& -0.11	\\
{2S}	& 3.69 +- .20		& 0.01\\
{1D}	& 3.80 +- .21		& $\!\!\!\!\mbox{---\phantom{oii}}$	\\
{2P}	& 4.04 +- .22		& 0.12	\\
{3S}	& 4.28 +- .24		& $\!\!\!\!\mbox{---\phantom{oii}}$\\
{2D}	& 4.36 +- .24		& $\!\!\!\!\mbox{---\phantom{oii}}$\\
{3P}	& 4.58 +- .26		& $\!\!\!\!\mbox{---\phantom{oii}}$\\
{4S}	& 4.79 +- .27		& $\!\!\!\!\mbox{---\phantom{oii}}$\\	\hline \hline
\end{tabular}
\end{table}

\begin{table}
\caption{Results for the bottomonium eigenstates using $V_{LGP+L}$ and the 
Cornell potentials in a constrained fit (see text).
Long dashes represent states that have not been observed experimentally.}
\label{tb:results_simultaneous_fit_bottom}
\centering
\begin{tabular}{cS[table-format=2.5]S[table-format=2.5]S[table-format=2.5]S[table-format=2.5]}
\hline \hline
\multicolumn{3}{c}{Bottomonium Spectrum} \\ \hline
\multicolumn{3}{c}{{$V_{LGP+L}$}}	\\ \hline
{State}	& {Mass}	& {Deviation from average}		\\
	& {(\si{\giga \electronvolt})}	& {spin state (\si{\giga \electronvolt})}	\\ \hline
{1S}	& 9.47  +- .30		& 0.04 \\
{1P}	& 9.86	+- .31		& -0.03\\
{2S}	& 10.00 +- .32		& -0.01	\\
{1D}	& 10.11 +- 1.6		& -0.05\\
{2P}	& 10.24 +- .33		& -0.01\\
{3S}	& 10.37 +- .33		& 0.01\\
{2D}	& 10.44 +- 1.4		& $\!\!\!\!\mbox{---\phantom{oii}}$\\
{3P}	& 10.56 +- .34		& 0.03\\
{4S}	& 10.67 +- .34		& 0.10\\
{3D}	& 10.73 +- 2.2		& $\!\!\!\!\mbox{---\phantom{oii}}$\\
{4P}	& 10.84 +- .35		& $\!\!\!\!\mbox{---\phantom{oii}}$\\ \hline
\multicolumn{3}{c}{{Cornell Potential}} \\ \hline
{State}	& {Mass}	& {Deviation from average}		\\
	& {(\si{\giga \electronvolt})}	& {spin state (\si{\giga \electronvolt})}	\\ \hline
{1S}	& 9.49 +- .31		& 0.06 \\
{1P}	& 9.84  +- .33		& -0.04\\
{2S}	& 10.00 +- .33		& -0.01\\
{1D}	& 10.10 +- .34		& -0.06	\\
{2P}	& 10.25 +- .34		& 0.00\\\
{3S}	& 10.39 +- .35		& 0.03\\
{2D}	& 10.45 +- .35		& $\!\!\!\!\mbox{---\phantom{oii}}$\\
{3P}	& 10.59 +- .36		& 0.05\\
{4S}	& 10.72 +- .37		& 0.14\\
{3D}	& 10.77 +- 2.4 		& $\!\!\!\!\mbox{---\phantom{oii}}$\\
{4P}	& 10.89 +- .38		& $\!\!\!\!\mbox{---\phantom{oii}}$\\ \hline \hline
\end{tabular}
\end{table}

The fact that the calculated spectra are very similar in the
$V_{LGP+L}$ and Cornell potential cases can be understood if we
note that, although the pure-OGE potentials were visibly different
(see Fig.\ \ref{fig:Comparison_Potentials}), 
the inclusion of the linear term brings the two
potentials closer, as shown in Fig.\ \ref{fig:full_pot_comparison}.

\vskip 3mm
An advantage of our approach is that we have direct access to the radial wave function $f(r)$. 
We plot, as an example, the wave 
functions\footnote{The wave functions obtained using our code are 
not normalized. We interpolate the data and normalize $f(r)$ a posteriori.} 
for the 1S state for both potentials in the 
charmonium and bottomonium cases in Fig.\ \ref{fig:wave function}. 
Thus, we see that the similarity between the two potentials (and the
obtained spectra) is present for the wave functions as well. Also, note 
that the wave function is more extended for the charmonium states,
as expected.

This direct access to the wave function can be of interest in other applications, such as
effective field theories, for which one needs information on the typical distance between the 
quarks \cite{Brambilla:2014jmp}. We estimate this quantity by computing
\begin{equation}
d = \int_0^\infty{r f(r)^2\, dr}\,. 
\label{eq:Typical_distance}
\end{equation}
Some of these typical distances are presented in Tables 
\ref{tb:Typical_distances_charm} and \ref{tb:Typical_distances_bottom}.

Finally, we could also estimate decay widths, which are proportional to 
$|R(0)|^2$. Notice, however, that this calculation would require a more 
strict control of the numerical integration in the region near 
the origin, since the function $R(r) = f(r)/r$ typically shows a 
divergence for $r\to 0$. This is beyond the scope of the present work.

\section{Conclusions}
\label{sec:Conclusions}

We briefly reviewed the potential-model approach for determining the 
spectrum of quarkonia and discussed the simplest such approach, the 
Cornell potential. We then modified the procedure for obtaining the
OGE potential, by replacing the free gluon 
propagator with one obtained using lattice simulations. The resulting 
$V_{LGP}$ potential is different from
the Coulomb-like potential, but is still non-confining. Inspection of Fig.\ \ref{fig:Comparison_Potentials}
shows that, up to the hadronic scale, the potential rises above zero, with 
a trend to rise further. This is no longer true for larger values of $r$, for which the potential
is damped.
In fact, in order to obtain a confining (linear) potential, the gluon propagator
should show a strong divergence, of $1/k^4$, in the infrared limit, as proven
in Ref.\ \cite{West:1982bt}.
Also, an oscillating behavior --- due to the complex poles of the lattice 
propagator \cite{Cucchieri:2011ig} --- is observed. 
We solve the associated Schr\"odinger equation numerically and compare our 
results with the spin-averaged spectrum in Tables
\ref{tb:input_data_charm} and \ref{tb:input_data_bottom}.
The spectrum obtained from this potential shows the interesting qualitative 
feature of approximately reproducing a few low-lying eigenstates. This confirms
our expectation that the short-distance behavior of the potential is improved
by using the fully nonperturbative gluon propagator instead of the tree-level
perturbative one. A quantitative description of the spectrum including
higher states is, however, not possible.

\begin{figure}
\centering
\caption{Comparison of the $V_{LGP+L}$ and Cornell Potentials.
For the value of the strong coupling constant $\alpha_s$, we choose the 
one used in the description of the charmonium spectrum (see Table 
\ref{tb:Input_Parameters}). The string tension is obtained from the
constrained fit (see Table \ref{tb:parameters_results_simultaneous_fit}).}
 \label{fig:full_pot_comparison}
	\vskip 1mm
	\includegraphics{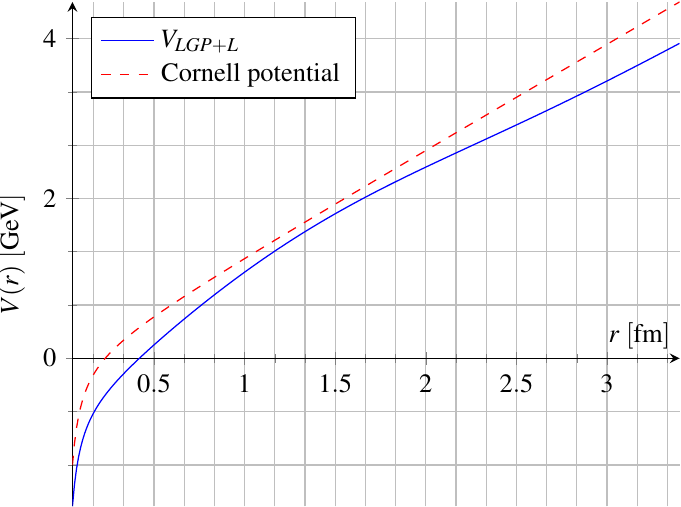}
\end{figure}

We therefore add a linear term to $V_{LGP}$, obtaining the 
$V_{LGP+L}$ potential in Eq.\ (\ref{eq:lgp+l_potential}).
We then compute the eigenenergies for the $V_{LGP+L}$ and Cornell 
potentials, both for charmonium and for bottomonium states.
The spectra obtained using $V_{LGP+L}$ show a slight
improvement over the Cornell potential, but no qualitative differences 
are observed. 
In particular, the resulting potentials are rather similar, as seen 
in Fig.\ \ref{fig:full_pot_comparison}.

\begin{figure}
\centering
\caption{Comparison of the wave function $f(r)$ of the 1S state for 
bottomonium and charmonium using the $V_{LGP+L}$ and Cornell Potentials.}
 \label{fig:wave function}
	\includegraphics{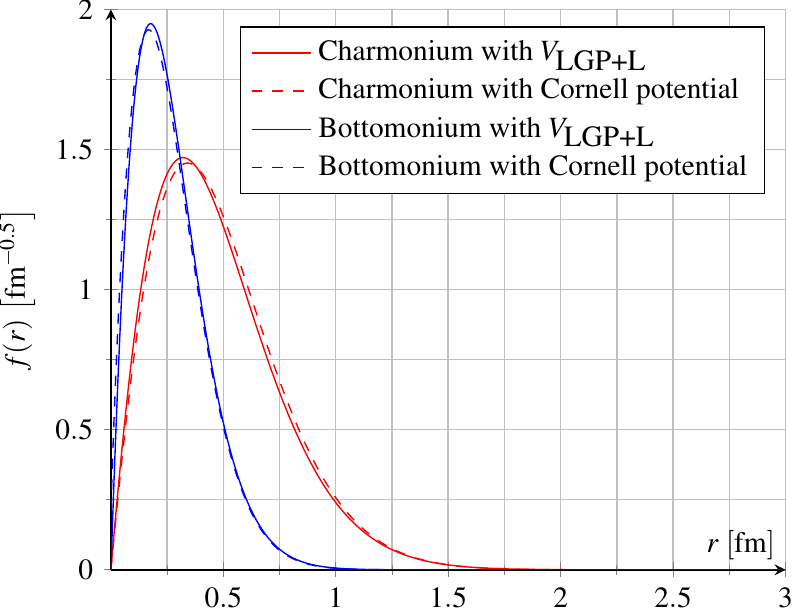}
\end{figure}

We were also able to obtain the wave functions for all the states, 
which allows us to estimate the corresponding interquark distances.
Let us note that the wave functions are remarkably similar for the 
$V_{LGP+L}$ and Cornell potentials (see Fig.\ \ref{fig:wave function}), 
even though the potentials are not identical
(see Fig.\ \ref{fig:full_pot_comparison}). This might suggest that 
the wave function is somewhat insensitive to details of the potential.
In fact, a visual comparison between our wave functions and the one 
presented in \cite[Fig.\ 5]{Kawanai:2013aca} (corresponding to a different 
parametrization of the Cornell potential)
shows that they are also essentially identical.

Let us mention that a study using a similar method was carried out in
Refs.\ \cite{Gonzalez:2011zc, Gonzalez:2012hx} to propose a potential 
for heavy-quarkonium states. In that case, the gluon propagator was taken from
a study of Schwinger-Dyson equations \cite{Aguilar:2008xm}.
This propagator is in qualitative agreement with the lattice results we
use. The main difference with respect to our study is that these authors
do not include the linear term in the potential, but consider an additive
contribution\footnote{Let us recall that a constant term in the interquark potential can 
also be related to the infrared divergence of the Fourier integral of a 
``confining'' gluon propagator $1/k^4$ \cite{Lucha:1991vn}.}
to the OGE potential, in such a way that the 
zero of the proposed potential coincides with the Cornell one.
This corresponds to fixing the (infinite) self-energy of the static sources 
\cite{Necco:2003jf}, which, however, is not present when considering
only the OGE diagram at tree level.
The spectrum obtained in \cite{Gonzalez:2012hx} is in general agreement 
with the expected values. 


\vskip 3mm
We stress again that the aim of this work was to gain a qualitative understanding of the 
interplay between perturbative and nonperturbative features of the interquark 
potential. As verified in our study, even though the full nonperturbative 
gluon propagator was used, the potential is non-confining, i.e.\ confinement 
is washed away by the use of the (tree-level) perturbative approximation 
for the interaction. 
Nevertheless, the resulting potential (with the 
addition of a linear term) provides a slightly better description of 
the spectra, with the same number of fit parameters as the Cornell potential.
Our preferred fit is done considering simultaneously the charmonium and 
bottomonium spectra, leaving as free parameters the two heavy quark masses
and the string tension $\sigma$.
We remark that leaving a known quantity as a free parameter
allows a further check of the model's consistency. 
Of course, it would be interesting to check if the inclusion of relativistic 
corrections, as done in Refs.\ \cite{Radford:2007vd,Olsson:1994cv},
would allow a more accurate description of the spectra.
 
\begin{table}[t]
\caption{Typical interquark distances for charmonium. Errors are expected to be negligible.}
\label{tb:Typical_distances_charm}
\centering
\begin{tabular}{cS[table-format=1.2]S[table-format=1.2]}
\hline
\multicolumn{3}{c}{Charmonium} \\ \hline \hline
{State}	& {$V_{LGP+L}$} 			& {Cornell Potential} \\ 
	& {distance (\si{\femto \metre})}	& {distance (\si{\femto \metre})} \\ \hline
{1S}	& 0.40					& 0.42\\
{1P}	& 0.64					& 0.64\\
{2S}	& 0.79 					& 0.78\\
{1D}    & 0.84                                  & 0.82\\
{2P}	& 0.97					& 0.95\\
{3S}	& 1.10 					& 1.07\\
{2D}    & 1.13                                  & 1.09\\
{3P}	& 1.25					& 1.21\\
{4S}    & 1.36                                  & 1.33\\ \hline \hline
\end{tabular}
\end{table}
\begin{table}[h]
\caption{Typical interquark distances for bottomonium. Errors are expected to be negligible.}
\label{tb:Typical_distances_bottom}
\centering
\begin{tabular}{cS[table-format=1.2]S[table-format=1.2]}
\hline \hline
\multicolumn{3}{c}{Bottomonium} \\ \hline
{State}	& {$V_{LGP+L}$} 			& {Cornell Potential} \\ 
	& {distance (\si{\femto \metre})}	& {distance (\si{\femto \metre})} \\ \hline
{1S}	& 0.22					& 0.24\\
{1P}	& 0.38 					& 0.38\\
{2S}	& 0.47					& 0.47\\
{1D}    & 0.51                                  & 0.50\\
{2P}	& 0.59					& 0.58\\
{3S}	& 0.67					& 0.65\\
{2D}    & 0.69                                  & 0.67\\
{3P}	& 0.77 					& 0.75\\
{4S}    & 0.84                                  & 0.81\\
{3D}    & 0.85                                  & 0.81\\
{4P}    & 0.93                                  & 0.90\\ \hline \hline
\end{tabular}
\end{table}

\begin{acknowledgments}
The authors thank B. Blossier and F.\ Navarra for useful comments. W.S.\ thanks the Brazilian funding agencies
FAPESP and CNPq for financial support. A.C.\ and T.M.\ thank CNPq for partial support.
\end{acknowledgments}

\end{document}